# Validation of musculoskeletal segmentation model with uncertainty estimation for bone and muscle assessment in hip-to-knee clinical CT images

**Mazen Soufi[1,*], Yoshito Otake[1,*], Makoto Iwasa[2], Keisuke Uemura[2], Tomoki Hakotani[1], Masahiro Hashimoto[3], Yoshitake Yamada[3], Minoru Yamada[3], Yoichi Yokoyama[3], Masahiro Jinzaki[3], Suzushi Kusano[4], Masaki Takao[5], Seiji Okada[2], Nobuhiko Sugano[6], and Yoshinobu Sato[1,*]**

[1]Division of Information Science, Graduate School of Science and Technology, Nara Institute of Science and Technology, 8916-5 Takayama-cho, Ikoma, Nara 630-0192, Japan
[2]Department of Orthopedic Surgery, Graduate School of Medicine, Osaka University, 2-2 Yamadaoka, Suita, Osaka 565-0871, Japan
[3]Department of Radiology, Keio University School of Medicine, 35 Shinanomachi, Shinjuku-ku, Tokyo 160-8582, Japan
[4]Hitachi Health Care Center, Hitachi Ltd., 4-3-16 Ose, Hitachi 307-0076, Japan
[5]Department of Bone and Joint Surgery, Graduate School of Medicine, Ehime University, Shitsukawa, Toon, Ehime 791-0295, Japan
[6]Department of Orthopaedic Medical Engineering, Graduate School of Medicine, Osaka University, 2-2 Yamadaoka, Suita, Osaka 565-0871, Japan
[1]msoufi,otake,yoshi@is.naist.jp

## ABSTRACT

Deep learning-based image segmentation has allowed for the fully automated, accurate, and rapid analysis of musculoskeletal (MSK) structures from medical images. However, current approaches were either applied only to 2D cross-sectional images, addressed few structures, or were validated on small datasets, which limit the application in large-scale databases. This study aimed to validate an improved deep learning model for volumetric MSK segmentation of the hip and thigh with uncertainty estimation from clinical computed tomography (CT) images. Databases of CT images from multiple manufacturers/scanners, disease status, and patient positioning were used. The segmentation accuracy, and accuracy in estimating the structures volume and density, i.e., mean HU, were evaluated. An approach for segmentation failure detection based on predictive uncertainty was also investigated. The model has shown an overall improvement with respect to all segmentation accuracy and structure volume/density evaluation metrics. The predictive uncertainty yielded large areas under the receiver operating characteristic (AUROC) curves (AUROCs≥.95) in detecting inaccurate and failed segmentations. The high segmentation and muscle volume/density estimation accuracy, along with the high accuracy in failure detection based on the predictive uncertainty, exhibited the model's reliability for analyzing individual MSK structures in large-scale CT databases.

## 1 Introduction

The advent of deep learning (DL)-based image segmentation has allowed for the fully automated, accurate, and rapid analysis of MSK structures from medical images[1–10]. These models assist in extracting the structure's shape and estimating diagnostic image biomarkers, such as volume and muscle density, for assessing muscle atrophy and fatty degeneration[11,12]. The models were tested on the rotator cuff[8], chest[3], hip and thigh[7], and abdominal muscles[1,2,4,9,10,13,14]. However, multiple issues exist in those studies that limit the reliable application in large-scale databases.

- Several studies addressed the segmentation of the muscles in only a single or a few 2D CT slices[1,15,16], which do not reflect the 3D properties of the muscles and depend on the subjective selection of the slices[17].
- The 3D muscle segmentation has also been attempted[8,10,18]; however, only a few muscles were addressed. A recent study[19] addressed the 3D segmentation of 27 hip and thigh muscles in CT images; however, the model was tested on a small database consisting of 12 cases, and the average accuracy was lower than that reported in a previous study targeting similar muscles[7].
- Current muscle segmentation approaches assess the model's accuracy only in cross-sectional area or volume estimation.

However, muscle density, which can be quantified based on the mean Hounsfield units (HU) in the CT image[20], has shown higher correlations with muscle strength and functions[11,21]. This necessitates the accuracy assessment of muscle density estimation, as well.

- Even though some studies attempted the analysis of large-scale databases[18,22], no rigid criteria were applied for segmentation failure detection. In other words, it is not clear how to determine whether the automatic predictions can be safely adopted, possibly corrected with moderate efforts, or better excluded for a reliable downstream analysis.

Our group has developed a segmentation tool, i.e., Bayesian UNet, that outputs the model's uncertainty, a.k.a *predictive uncertainty* in addition to the target segmentations[7]. The model was validated on a database of 20 cases of hip osteoarthritis (hip OA) patients. It has shown high accuracy in segmenting 19 hip and thigh muscles as well as the possibility of predicting the segmentation accuracy in *unannotated* CT images based on the predictive uncertainty. In the future, we want to leverage this tool to segment large-scale databases of CT images collected from many health centers[23], and analyze the impact of the demographic and disease factors in the Japanese population. These databases include large variations from the training data, such as manufacturer/scanner, imaging conditions, and disease variations, which may lead to segmentation failure due to the domain shift problem[24–27]. Fitzpatrick et al. reported the automated volumetric and demographic analysis of the iliopsoas muscle segmented from magnetic resonance (MR) images of 5,000 subjects[22]. Their database was collected from the UK Biobank database[28], which, in contrast to ours, has a unified imaging scanner and protocol that mitigates the domain shift problem. The predictive uncertainty was also addressed in previous studies to predict the segmentation accuracy in *unannotated* images[7,29–32]; however, the analysis was limited to small databases, and no quantitative criteria were applied for the detection of the segmentation failures for the down-stream analysis.

In this study, we report the preparations conducted to employ the model for muscle segmentation in the large-scale database. In particular, a larger fully annotated database consisting of 50 cases of hip OA patients acquired by two CT scanners has been prepared. In addition, the model's capacity has been increased to account for the enlarged training database. The major contributions of this work are as follows:

- Investigating the segmentation accuracy and volume/intensity prediction in 22 MSK structures from four databases of CT images acquired from multiple manufacturers/scanners with various disease conditions, and patient positioning, i.e., standing and supine positions.
- Assessing the accuracy of the predictive uncertainty as a predictor of the segmentation accuracy under various imaging conditions and disease variations and suggesting quantitative criteria for detecting segmentation failures.
- Showcasing the capability of the predictive uncertainty and suggested criteria in detecting segmentation failures at a large database of >2,500 volumetric CT images of hip OA patients.

To the best of our knowledge, this is the first study to investigate the different practical aspects facing DL-based segmentation models towards the reliable analysis of MSK structures in large-scale CT databases.

## 2 Materials and Methods

### 2.1 CT images and annotations

In this study, databases of CT images from multiple manufacturers/scanners, disease status, and patient positioning were used. Table 1 summarizes the characteristics of the databases (DBs) used in this study. DB#1 included images from 50 unilateral HOA patients (mean age: 61.4 ± 13.0 yrs, min: 30 yrs, max: 86 yrs; 44 females, 6 males) acquired by two scanners from different generations by the same manufacturer (HiSpeed "old" (N=20) and Optima CT660 "new" (N=30), GE Healthcare, Milwaukee, WI). The images were resampled so the slice interval became 1.0 mm throughout the entire volume. The disease severity was assessed using Crowe[33] and Kellgren and Lawrence (KL)[34] grading, in which higher grades indicate higher disease severity. The affected sides were those with KL,Crowe>1.

The three databases DB#2-4 were for subjects without HOA. DB#2 was collected from a public database[35], including 18 cases (age anonymized; 13 females, 5 males) with soft tissue sarcoma acquired by a scanner from the same manufacturer as DB#1 but a different model (Discovery ST, GE Healthcare, Milwaukee, WI). DB#3 included images for 10 subjects (mean age: 50.1 ± 7.6 yrs, min: 41 yrs, max: 64 yrs; 10 males) who were scanned for the diagnosis of colorectal cancer using a scanner from a different manufacturer (Supria, Hitachi Medical, Tokyo, Japan). DB#4 included images of 20 healthy volunteers (mean age: 65.1 ± 6.3 yrs, min: 55 yrs, max: 76). The images were acquired for the volunteers in the supine and standing positions. The supine images were acquired with a 320-row detector CT scanner (Aquilion ONE, Canon Medical Systems Corporation, Otawara, Japan), while the standing images were acquired with an upright 320-row detector CT (prototype TSX-401R, Canon Medical Systems Corporation, Otawara, Japan)[36].

DB#5 included images for 2579 uni/bilateral HOA patients (mean age: 61.8 ± 15.2 yrs, min: 13 yrs, max: 98 yrs; 2062 females, 497 males) collected from the same institution as DB#1. The affected and unaffected sides were assigned based on an automatic grading model with an accuracy of .962 (for details about the automated grading model, see Ref.[37]).

**Table 1.** CT image characteristics.

| Database | Inst. | Diagnosis | Patient positioning | No. of cases | Modality | Matrix size | In-plane resolution [mm] | Slice interval [mm] |
|---|---|---|---|---|---|---|---|---|
| Training and testing (5-fold cross-validation) | | | | | | | | |
| DB#1 | Osaka Univ. Hosp. | Unilateral HOA | Supine | 20 | HiSpeed, GE | $512^2$ | $0.703^2$ – $0.742^2$ | 1.0-6.0* |
| | | | | 30 | Optima CT660,GE | $512^2$ | $0.703^2$ – $0.820^2$ | 1.25 |
| External validation (small-scale, with GMed ground-truth labels) | | | | | | | | |
| DB#2 | TCIA | Soft tissue sarcoma | Supine | 18 | Discovery ST,GE | $512^2$ | $0.977^2$ | 3.75 |
| DB#3 | Hitachi Medical Care Center | Colorectal cancer | Supine | 10 | Supria, Hitachi | $512^2$ | $0.685^2$ | 0.63 |
| DB#4 | Keio Univ. Hosp. | Normal | Supine | 20 | Aquilion ONE, Canon Medical Systems | $512^2$ | $0.683^2$ | 0.5 |
| | | | Standing | 20 | prototype TSX-401R, Canon Medical Systems | $512^2$ | $0.683^2$ | 0.5 |
| External validation (large-scale, without ground-truth labels) | | | | | | | | |
| DB#5 | Osaka Univ. Hosp. | Uni/ bilateral HOA | Supine | 460 | HiSpeed, GE | $512^2$ | $0.703^2$ – $0.742^2$ | 1.0-6.0* |
| | | | | 2119 | Optima CT660,GE | $512^2$ | $0.703^2$ – $0.820^2$ | 0.675-3.75 |

*pelvis and proximal femur: 2.0 mm, femoral shaft region: 6.0 mm, distal femur region: 1.0 mm .

A collaborative group consisting of a health science researcher with a medical physics background, computer science researchers, and orthopedic surgeons specializing in musculoskeletal imaging created and validated the ground-truth (GT) labels of 19 muscles and three bones (see Fig. 2) in DB#1 and GMed in DBs#2-4. The annotations of the 50 cases in DB#1 passed through multiple annotation and validation cycles. The annotations were first created using a pre-trained model[7], and the automated segmentations were corrected using 3D Slicer[38]. DB#1 was used to train and validate the segmentation models on all structures, while DBs#2-5 were used for external validation on small-scale (DBs#2-4) and large-scale (DB#5) databases.

## 2.2 Overall scheme

Figure 1 shows the overall scheme of validating the segmentation model for the automated assessment of bones and muscles in CT images. The CT image was input to the model, where each axial slice was processed to segment the bones and muscles. Each bone and muscle was extracted from the concatenated volume of all slices for qualitative, i.e., muscle density visualization, and quantitative, i.e., volume and mean HU assessments. Besides the bone/muscle labels, the structure-wise predictive uncertainty was computed based on Monte-Carlo (MC) dropout sampling[7,39].

## 2.3 Image segmentation

In this study, a cascaded 2D Bayesian U-Net model, which outputs the predicted structure labels with pixel-wise predictive uncertainty maps, was used[7]. The baseline model consisted of an encoder and decoder composed of multiple down/upsampling layers (hereinafter called *layers* for simplicity). Further details on the baseline model architecture can be found in Ref.[7]. Two modifications were made to the baseline architecture: 1) increasing the depth of the model (i.e., using six instead of five encoder layers), and 2) adding a batch normalization layer[40] to the basic convolutional blocks, which stabilizes the training of large neural network models and improves the overall performance[29]. Similar to[7], at the inference time, the mean and the variance of 10 MC dropout samples were used to obtain the output label and voxel-wise uncertainty map, respectively. The structure-wise predictive uncertainty was computed as the average of the voxel-wise uncertainty map within the segmented label.

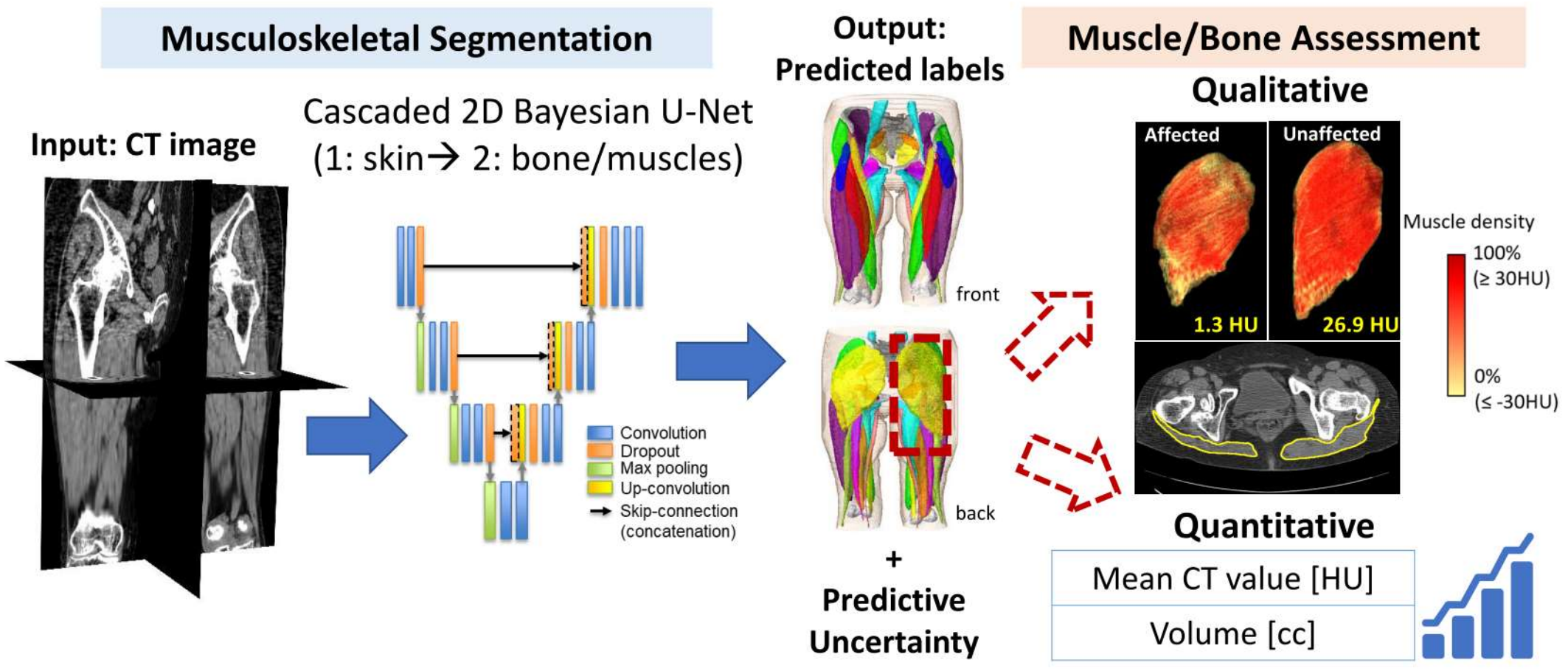

**Figure 1.** Overall scheme for validation of musculoskeletal segmentation model for automated assessment of bones and muscles in CT images with uncertainty estimation.

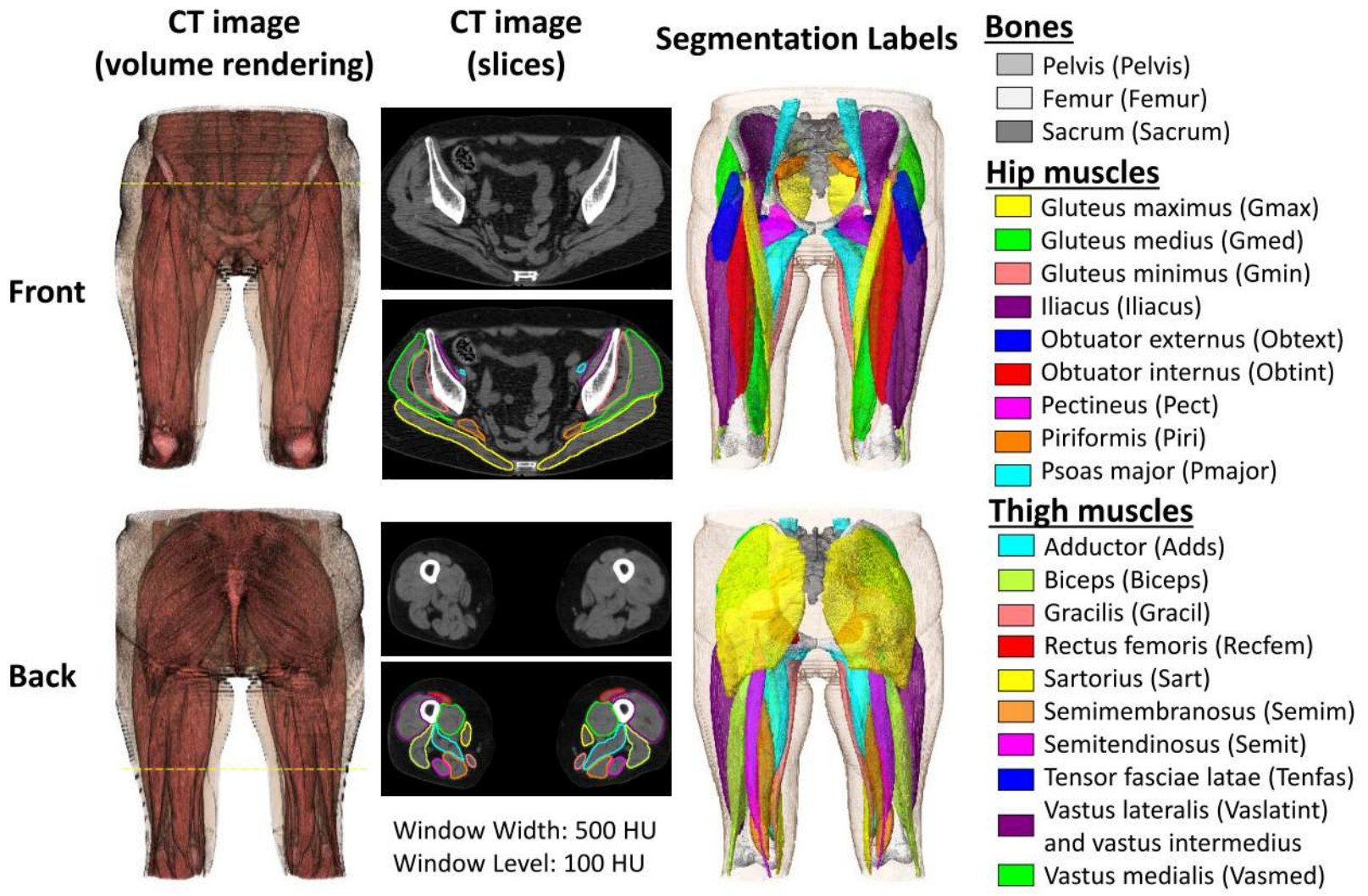

**Figure 2.** Segmentation labels of the bones and muscles. Abbreviations are written within brackets.

To investigate the impact of the modified model and larger annotated database, including 50 cases, the performance was compared with the baseline model consisting of five encoder layers and validated on 20 or 50 cases. For simplicity, the aforementioned models were termed (*5layers,20*), (*5layers,50*), and (*6layers,50*). The parameters were ~10 M and ~44 M for the *5layers* and *6layers* models, respectively.

### 2.4 Muscle/bone assessment

The labels predicted by the segmentation model were used to assess each structure's volume and muscle density. The volume was computed as a multiplication of the number of voxels by the size of each voxel in centimeter cubes (cc) normalized by the subject's height. The muscle density was computed as the mean of the intensity or CT values in HU within the segmented label. Structures on the right and left sides of the patient's body were assessed separately based on a postprocessing using connected component analysis (CCA). An additional watershed algorithm and CCA were used to separate the right and left sides when connected (e.g., the connection of right and left hemi-pelvises at the pubic symphysis).

A transfer function was used to comprehensively convert the HU values into scalar muscle density to visualize lean muscle and intramuscular fat. HU values less than -30 HU were considered fat, values within the range $[-30,30]$ were considered muscle/fat composite, and values larger than 30 HU were considered lean muscle[20]. Color and opacity transfer functions were used to visualize the transformed image (see Fig. 5, right).

### 2.5 Evaluation metrics

The segmentation accuracy, and accuracy in estimating the structures volume and density, i.e., mean HU, were evaluated. The segmentation accuracy was evaluated using the Dice coefficient (DC) and average symmetric surface distance (ASD). DC assesses the overlap between the GT and predicted labels. ASD assesses the surface distance, i.e., surface error, to assess the presence of small yet distant false positive structures. The predicted volume and mean HU accuracy were evaluated using the absolute difference between the quantities measured at the GT and predicted labels. The volume error (average volume error [AVE]) was computed as a percentage relative to the GT volume. The intensity error (average intensity error [AIE]) was reported as the average of absolute differences between the mean HUs of the GT and predicted labels.

The accuracy of the predictive uncertainty for detecting inaccurate (correctable with moderate human effort) or failed (correctable with notable efforts) segmentations was investigated. For each structure, a threshold based on the standard deviation of DC was determined to consider the segmentation inaccurate or failed. To make the threshold setting more statistically robust against outliers, the median absolute deviation (MAD) was used. Particularly, a threshold of $Median_{DC} + 1.4826 * k * MAD_{DC}$) was used, where k was set as -2 or -3 for inaccurate or failed segmentations, respectively. The area under the receiver operating characteristic (AUROC) curves of the predictive uncertainty based on the DC threshold was used to assess the detection accuracy. The AUROCs of the predictive uncertainty from both *5layers,20* and *6layers,50* were computed. Linear regression lines were computed between DC (dependent) and the predictive uncertainty (independent) for each structure and the averages of all structures combined.

### 2.6 Statistical analysis

The concordance correlation coefficient ($CCC$)[41] was used to assess the agreement between the GT and predicted volume and mean HU. The Pearson correlation coefficient ($\rho$) assessed the linear relationship between the predictive uncertainty and DC. To investigate the statistical significance of the differences between paired measurements, the Shapiro test was first used to assess the normality of the different distributions. Student's t-test was used when normality was found. Otherwise, the Wilcoxon signed-rank test was used. A probability of $p = .05$ was considered significant in all tests. Bonferroni correction was used when multiple comparisons between the models were made.

### 2.7 Implementation details

The proposed approach was developed and validated in Python and Keras[42,43]. The segmentation models (*5layers,20*), (*5layers,50*), and (*6layers,50*) were trained and validated on DB#1 based on 5-fold cross-validation. For a matched comparison among the models, the remaining 30 out of the 50 cases used in training the (*5layers,20*) model were used in the inference phase. Models with 5 and 6 layers were retrained on all the images in DB#1 and were used to predict the labels in DBs#2-5. The quantitative validation of DBs#2-4 was limited to the Gmed muscle, whereas the average of the predictive uncertainty in all structures DB#5 were used. The predictive uncertainty thresholds used for failure detection in DB#5 were derived based on the linear regression lines computed in DB#1.

The segmentation model training and inference were performed on a Linux-based cluster of servers with graphical processing units (GPUs; Nvidia Corporation, Santa Clara, CA, USA). Similar to the previous study[7], the (*5layers,20*) model was trained for 150k iterations with a batch size of 3, whereas the models *5layers,50* and *6layers,50* were trained for 200k iterations due to the increased training data and model capacity. The inference time per volume (approximately 500 CT slices) models was approximately 3 minutes.

## 3 Results

### 3.1 Segmentation accuracy and predictive uncertainty

The improved model *6layers,50* has shown overall improvement with respect to all evaluation metrics. Figure 3 shows the segmentation accuracy, predictive uncertainty, and volume/mean HU accuracy of the three models. Each point represents the average metric value of all structures in a single subject. The accuracy of the *6layers,50* model was significantly higher than that of the *5layers,20* in terms of all metrics. The average DC of the *6layers,50* model was .945±.015, with an average increase of 1.2% at all structures compared with *5layers,20* (p<.017). An average improvement of approximately 0.4 mm was observed in ASD (p<.017). The improvement by *6layers,50* model was statistically significant in most MSK structures, as shown in Supplementary Figs. A.1 (DC), A.2 (ASD).

The box plots in Fig. 3(b) show that the uncertainty proportionally decreased with the improved segmentation accuracy in Fig. 3(a) regarding the number of cases and model depth. Scatter plots of DC versus the predictive uncertainty for each model are depicted in Supplementary Fig. A.6,(a). Linear relationships with strong correlations were obtained between the segmentation accuracy and the predictive uncertainty by all the models. A strong correlation of $\rho$=-.79 was obtained in the *6layers,50* model. This emphasizes the usability of the predictive uncertainty as a predictor of the segmentation accuracy, which supports the findings by the previous studies[7,29].

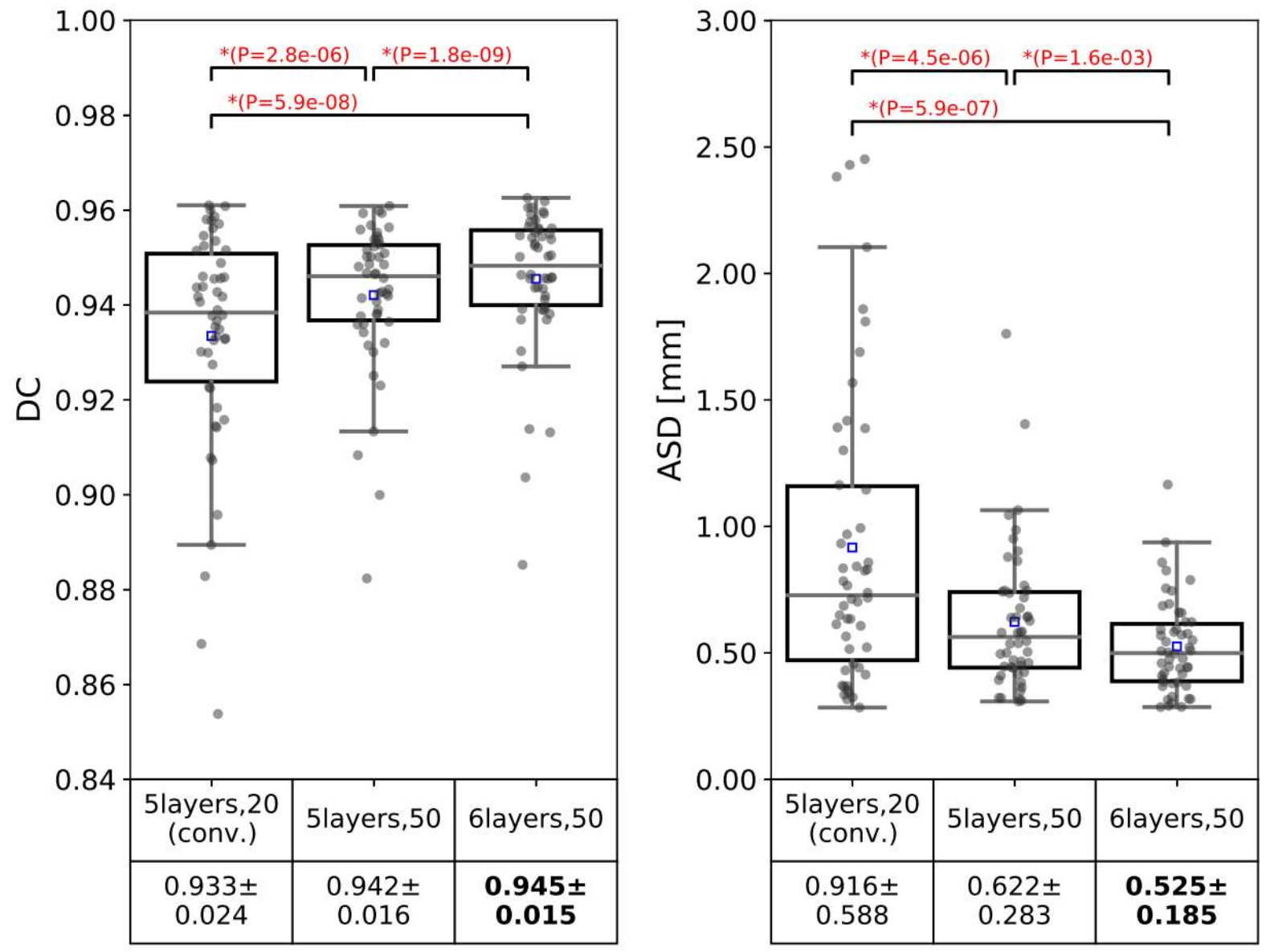

(a) Segmentation accuracy

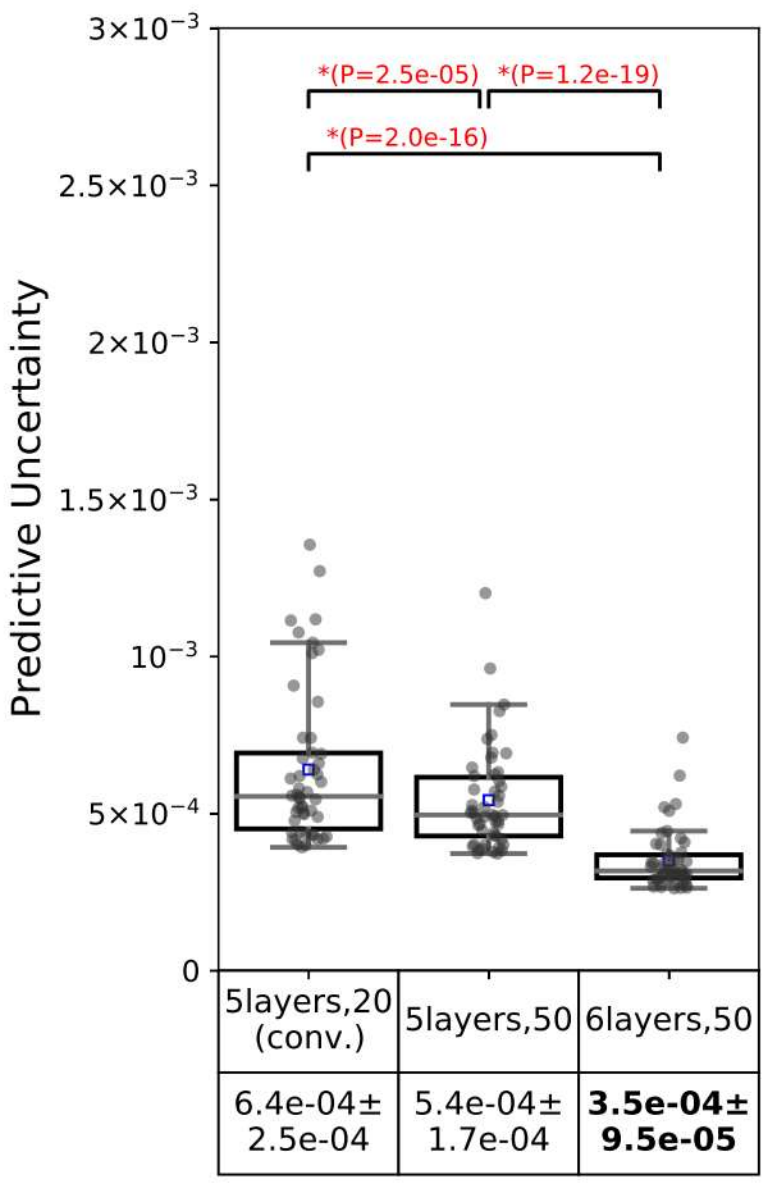

(b) Predictive uncertainty

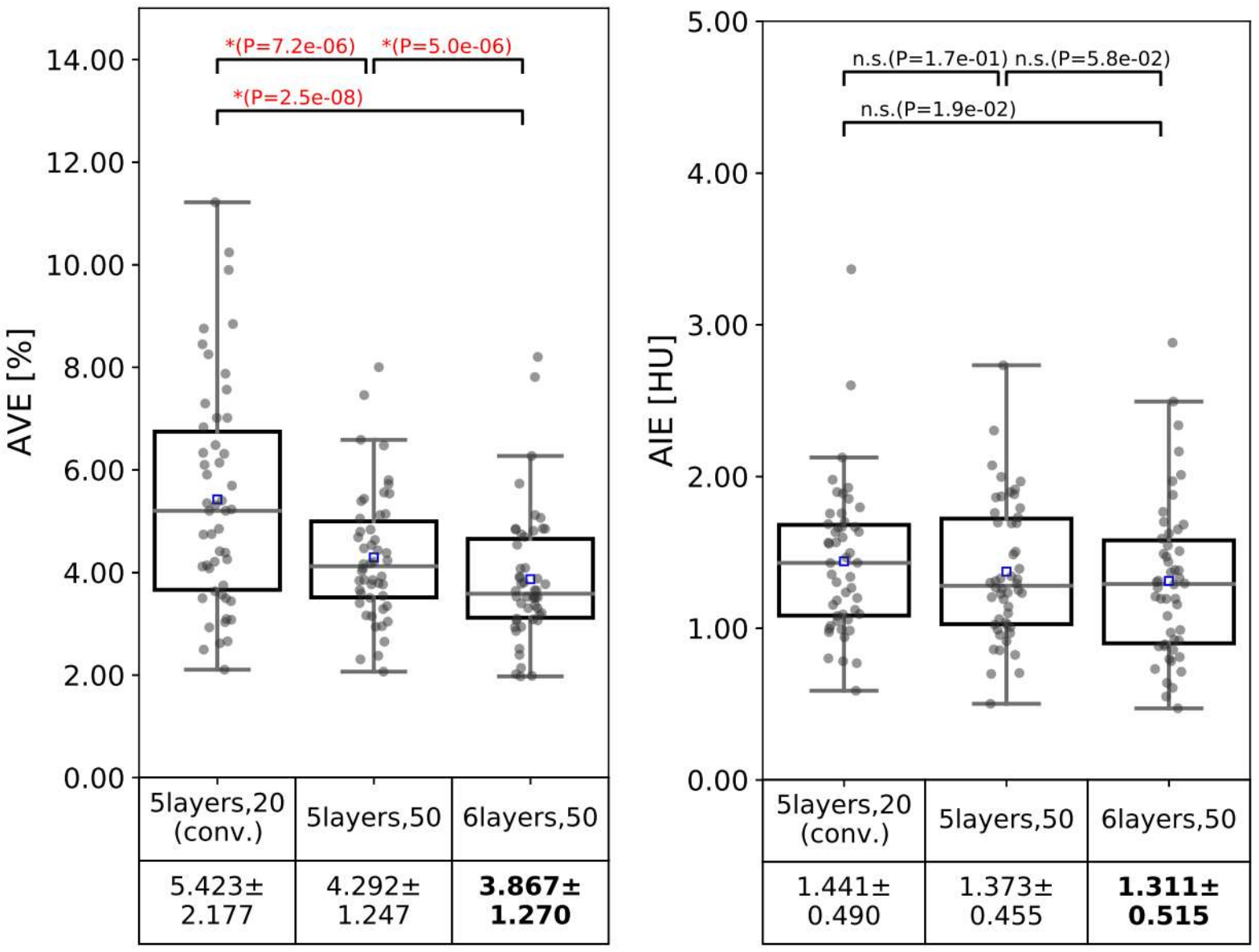

(c) Volume/mean HU accuracy

**Figure 3.** Distributions of the segmentation accuracy (a), predictive uncertainty (b), and volume/mean HU accuracy (c) of the bones and muscles (averaged on all structures) by each model. Horizontal lines in the boxes represent the medians, while blue boxes represent the means. Detailed values are depicted in Supplementary Figs. A.1-A.5. DC: Dice coefficient, ASD: average symmetric surface distance, AVE: average volume error, AIE: average intensity error, n.s.: not significant, *: p<.017, Student's t-test or Wilcoxon signed rank sum test with Bonferroni correction.

Figure 4 shows the relationship between the average DC and average predictive uncertainty of all structures in each patient in DB#1 and the corresponding ROC curve for failure detection. The predictive uncertainty of both models (*5layers,20* and *6layers,50*) yielded high AUROCs ($\geq$ .95) in detecting inaccurate and failed segmentations. Table 2 shows the AUROCs of each structure. The median AUROCs of all structures by *6layers,50* for detecting inaccurate and failed segmentations were .979 and .959, respectively. Obtext and Obtint had the lowest accuracy. Supplementary Fig. A.9 shows the detailed results of each structure. Supplementary Figure A.11 shows scatter plots of the predictive uncertainty by the two models in DB#5. Based on the thresholds computed in DB#1 (Fig. 4), three representative cases were visualized. The improved segmentation by the *6layers,50* model can be observed in the three cases and the scatter plot. The representative case of the failed segmentation exhibits unusual positioning of the hip, possibly due to the patient's discomfort as a result of the disease.

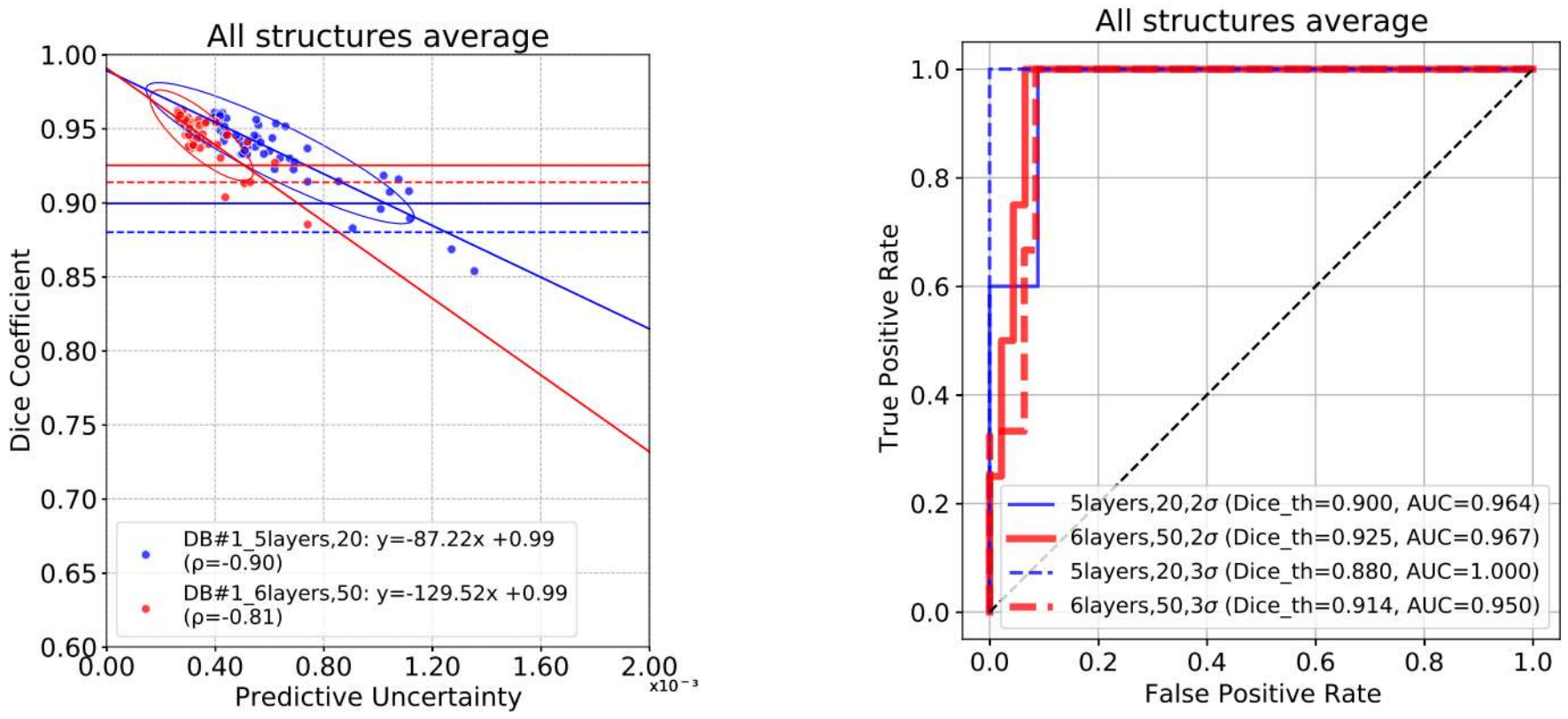

(a) Dice coefficient (DC) vs. Predictive Uncertainty (b) ROC Curves

**Figure 4.** Receiver operating characteristic (ROC) curves of the inaccurate and failed segmentation detection in DB#1 using the predictive uncertainty. Thresholds were determined based on the median absolute deviations ($\sigma$) of the DC.

### 3.2 Relationship between segmentation accuracy/predictive uncertainty and disease stage

Figure 5 shows the distributions of the evaluation metrics and predictive uncertainty at the internal DB#1 (a) and predictive uncertainty in the large-scale database DB#5 (b) in terms of the HOA disease status (unaffected vs. affected) in each body side. The model *6layers,50* showed statistically significant improvement ($p < .01$) in all the structure groups in DB#1. The proportional relationship between the accuracy and predictive uncertainty can also be observed in all structure groups, where smaller uncertainty was accompanied by increasing accuracy. The unaffected sides significantly showed higher accuracy in the bones than the affected ones in DB#1. All groups had a similar tendency in DB#5, where the affected sides had higher predictive uncertainty. In addition to the sensitivity to the variations in the positioning, as shown in Fig. A.11, this shows a possible impact of the disease status on the performance of the segmentation model in large-scale databases.

### 3.3 Validation on a multi-manufacturer/scanner database

Figure 6 shows the evaluation metrics and predictive uncertainty of Gmed muscle segmented at the databases DB#1-4 from multiple manufacturers/scanners and disease variations (see Table. 1). Representative cases ($5^{th}$ (▲) and $95^{th}$ (▼) quantiles of the predictive uncertainty visualized in Supplementary Figs.A.7 and A.8) are depicted. Statistically significant improvements in the DC, ASD, and AIE were observed in the four databases using the model *6layers,50*. The predictive uncertainty was obviously related to the accuracy metrics, where low uncertainty cases mostly had high accuracy metric values and vice versa. Table 3 summarizes the means and SDs of the evaluation metrics in the four databases. Using the *6layers,50* model, the segmentation accuracy at a predictive uncertainty of $5 \times 10^{-4}$ was > .90 (DC) and approximately less than 2.00 mm (ASD) in all the databases. Notably, a sub-HU accuracy was obtained in predicting the mean HU in the four databases. Overall improvements were observed in all the muscles by the *6layers,50* model (see Supplementary Fig. A.7). At Gmed, the *5layers,20* model failed to capture the boundaries with the Gmax muscle, which reduced its segmentation accuracy, whereas the errors were less in the *6layers,50* model results. The Psoas muscle's lower part was undersegmented in multiple instances by the *5layers,20* model. Noteworthy, only a slight degradation in accuracy was observed at standing compared with the supine positioning in DB#4.

Table 4 summarizes the predictive uncertainty and its correlations with the segmentation accuracy (DC) of Gmed in the four databases. With the four databases combined, both models yielded strong correlations, where the average PCCs for the *5layers,20* and *6layers,50* models were -.85 and -.60, respectively. Table 5 shows the AUROCs of failure detection at the four databases. The median AUROC by the *6layers,50* were .963 and .995 for inaccurate (-2$\sigma$) and failed (-3$\sigma$) segmentation detection, respectively.

**Table 2.** Area under receiver operator curve (AUROC) of the predictive uncertainty and segmentation accuracy (Dice coefficient; DC) of all structures DB#1. $\sigma$ indicates the threshold computed based on the median absolute deviation of DC and used for the detection of inaccurate (-2$\sigma$) and failed (-3$\sigma$) segmentations.

| Group | Structure | 5layers,20 -2$\sigma$ | 5layers,20 -3$\sigma$ | 6layers,50 -2$\sigma$ | 6layers,50 -3$\sigma$ |
|---|---|---|---|---|---|
| Hip muscles | Gmax | .935 | .970 | .874 | .958 |
| | Gmed | .989 | .990 | 1.000 | .990 |
| | Gmin | .864 | .802 | .959 | .959 |
| | Iliacus | .943 | – | .865 | .837 |
| | Obtext | .696 | .592 | .534 | .413 |
| | Obtint | .552 | – | .693 | – |
| | Pect | .962 | .990 | .965 | .980 |
| | Piri | .987 | 1.000 | 1.000 | 1.000 |
| | Pmajor | .991 | .979 | .963 | .916 |
| Thigh muscles | Adds | .936 | 1.000 | .894 | .990 |
| | Biceps | .977 | .967 | .898 | .898 |
| | Gracil | 1.000 | 1.000 | .913 | .996 |
| | Recfem | .981 | .996 | .965 | .970 |
| | Sart | .973 | .996 | .876 | .978 |
| | Semim | .924 | .969 | .938 | 1.000 |
| | Semit | .875 | .867 | .894 | 1.000 |
| | Tenfas | .964 | .908 | .819 | .682 |
| | Vaslatint | .920 | .911 | .951 | .940 |
| | Vasmed | .896 | .984 | .853 | .986 |
| Bones | Pelvis | .917 | .939 | .891 | .765 |
| | Femur | .995 | .995 | .904 | .948 |
| | Sacrum | .920 | .958 | .750 | .791 |
| All structures average | | .964 | 1.000 | .967 | .950 |
| Median | | .943 | .979 | .898 | .959 |

**Table 3.** Comparison between evaluation metrics of Gmed segmentation in four databases. p: p-value of the difference between 5layers and 6layers models (Student's t-test if normal distribution, Wilcoxon signed rank test otherwise, with Bonferroni correction), n.s.: not significant.

| | DC↑ | | | ASD [mm]↓ | | | AVE [%]↓ | | | AIE [HU]↓ | | |
|---|---|---|---|---|---|---|---|---|---|---|---|---|
| | 5layers,20 | 6layers,50 | p | 5layers,20 | 6layers,50 | p | 5layers,20 | 6layers,50 | p | 5layers,20 | 6layers,50 | p |
| DB#1 | 0.951±.015 | 0.961±.011 | * | 0.910±2.436 | 0.432±0.132 | * | 4.426±3.303 | 2.355±2.038 | * | 1.132±0.906 | 0.811±0.663 | n.s. |
| DB#2 | 0.915±.018 | 0.924±.015 | * | 2.195±3.359 | 1.811±2.029 | n.s. | 3.681±3.971 | 2.869±2.048 | n.s. | 1.915±1.685 | 0.959±0.665 | * |
| DB#3 | 0.879±0.049 | 0.938±0.012 | * | 2.090±0.807 | 0.985±0.255 | * | 8.500±9.491 | 3.033±1.950 | n.s. | 0.452±0.324 | 0.603±0.253 | n.s. |
| DB#4su | 0.955±.017 | 0.974±0.005 | * | 0.762±0.543 | 0.288±0.057 | * | 4.304±6.074 | 1.520±0.794 | * | 0.767±0.409 | 0.216±0.109 | * |
| DB#4st | 0.922±.048 | 0.969±0.005 | * | 0.985±0.667 | 0.314±0.058 | * | 6.195±7.852 | 1.347±0.842 | * | 0.640±0.364 | 0.198±0.102 | * |

### 3.4 Muscle/bone assessment

Table 6 compares the volume and mean HU prediction applied to the GT and auto segmentations in DB#1 obtained from the *6layers,50* model. The measurements of the unaffected and affected HOA sides were reported separately. Most structures exhibited substantial agreement between the GT and auto measurements on both sides ($\rho \geq .95$). The Piri muscle showed weak agreement in the volume and HU measurements, whereas the Tenfas muscle showed weak agreement only in mean HU. In both the volume and mean HU predictions, the unaffected side has shown a slightly larger MAE than the affected side. MAE of the predicted volumes at the bones and muscles for the affected and unaffected sides was 1.77±1.06 cc/$m^2$ and 1.89±1.17 cc/$m^2$, respectively. The MAE of the mean HU for the affected and unaffected sides was 1.46±0.95 HU and 1.38±0.89 HU, respectively. Notably, a sub-HU MAE was obtained at the Gmax, Gmed, Adds, Biceps, Recfem, Semim, Vaslatint, and Vasmed muscles on the affected and unaffected sides.

Figure 7 shows representative case (median DC) segmentations with muscle histograms and 3D volume rendering of muscle density of the GMax and GMed muscles. High reproducibility of the GT-based histograms and muscle density visualizations could be observed. In particular, the auto segmentations could comprehensively reproduce lean muscle (red) and fat (yellow) portions.

**Table 4.** Predictive uncertainty (mean±standard deviation ”std”) and correlation (Pearson correlation coefficient $\rho$) with Dice coefficient of the Gmed in four databases.

| Model | 5layers,20 | | 6layers,50 | |
|---|---|---|---|---|
| | mean±std ($\times 10^{-4}$) | $\rho$ | mean±std ($\times 10^{-4}$) | $\rho$ |
| DB#1 (N=50) | 4.601 ± 1.458 | -0.78 | 2.678 ± 0.593 | -0.72 |
| DB#2 (N=18) | 5.703 ± 1.703 | -0.77 | 3.367 ± 1.367 | -0.88 |
| DB#3 (N=10) | 14.142 ± 5.294 | -0.82 | 2.710 ± 0.523 | 0.12 |
| DB#4su (N=20) | 5.225 ± 2.675 | -0.93 | 2.603 ± 0.539 | -0.88 |
| DB#4st (N=20) | 8.774 ± 5.142 | -0.97 | 3.102 ± 0.851 | -0.88 |

**Table 5.** Area under receiver operator curve (AUROC) of the predictive uncertainty and segmentation accuracy (Dice coefficient; DC) of Gmed in the four databases. $\sigma$ indicates the threshold computed based on the median absolute deviation of DC and used for the detection of inaccurate (-2$\sigma$) and failed (-3$\sigma$) segmentations.

| Model | 5layers,20 | | 6layers,50 | |
|---|---|---|---|---|
| | -2$\sigma$ | -3$\sigma$ | -2$\sigma$ | -3$\sigma$ |
| DB#1 (N=50) | .989 | .990 | 1.000 | .990 |
| DB#2 (N=18) | .889 | .882 | 1.000 | – |
| DB#3 (N=10) | .813 | .813 | – | – |
| DB#4su (N=20) | 1.000 | 1.000 | 1.000 | 1.000 |
| DB#4st (N=20) | 1.000 | 1.000 | 1.000 | 1.000 |
| Median | .989 | .990 | 1.000 | 1.000 |

## 3.5 Impact of the number of training images and dropout samples

### 3.5.1 Number of training images

Table 7 shows the impact of the number of training images on the segmentation accuracy (DC, ASD) of the 6layers model applied to DB#1. The highest accuracy was obtained when 40 cases were used. However, no statistically significant differences were observed in comparison to 30 cases. Compared with cases fewer than 30 cases, statistically significant differences were observed. In addition, strong correlations between DC and the predictive uncertainty were obtained in all numbers of training cases (see Supplementary Fig.A.6,b). This emphasizes the generalizability of the predictive uncertainty as a predictor of the segmentation accuracy regardless of the number of training images.

### 3.5.2 Number of dropout samples

Table 8 shows the impact of the number of training samples on the segmentation accuracy by the 6layers model applied to DB#1. No improvement was observed by increasing the samples to larger than 10 samples. This indicates that 10 samples are sufficient to obtain a stable performance by the model.

# 4 Discussion

This study validated a DL model for segmentation of MSK structures with uncertainty estimation in clinical CT images. The novelty of this work is that it showed the usability of the predictive uncertainty for predicting the MSK segmentation accuracy and detecting segmentation failures in databases of CT images from multi-manufacturers/scanners and with disease and positioning variations, such as supine and standing, and with different scales, including a large-database with 2579 CTs. This showed the possibility of using the predictive uncertainty as a tool for detecting the failed segmentation in *unannotated* CT images. The study also exhibited the potential of the *6layers,50* model in producing accurate segmentations for assessing the muscle/bone volume and mean intensity, with DC >.90 in almost all the muscles and >.95 in the bones (see Supplementary Fig. A.1). The validation on the external databases has shown high generalizability of the model's performance, where a DC>.95 and an AIE<1 HU were obtained in evaluating the Gmed segmentations, and the predictive uncertainty could detect the cases with segmentation failures.

Systematic improvements were observed using the *6layers,50* model at all the structure groups, regardless of the disease status. However, the Piri muscle showed the smallest DC of .845±.091, with the largest ASD, AIE, and AVE (see Supplementary Figs. A.1, A.2, A.4, A.5). The degraded accuracy could be interpreted by the location of this muscle among various bony, abdominal, and vascular structures, making it challenging for automated segmentation. 3D segmentation models[44] might improve the segmentation accuracy of this muscle as they better involve the volumetric relationships with the surrounding structures.

**Table 6.** Comparison between affected and unaffected sides of the muscles and bones in DB#1 in terms of normalized volume and mean HU using ground truth (GT) and auto (Auto) segmented labels. CCC: Concordance correlation coefficient between GT and predicted measurements, MAE: mean absolute error between GT and predicted measurements

| | | Normalized Volume [cc/m$^2$] | | | | | | | Mean HU [HU] | | | | | | |
|---|---|---|---|---|---|---|---|---|---|---|---|---|---|---|---|
| | | Affected | | | Unaffected | | | | Affected | | | Unaffected | | | |
| Group | Structure | GT | Auto | MAE | GT | Auto | MAE | CCC | GT | Auto | MAE | GT | Auto | MAE | CCC |
| Hip muscles | Gmax | 230.27 | 230.90 | 3.65 | 255.38 | 255.56 | 4.05 | 1.00 | 19.33 | 19.33 | 0.52 | 25.03 | 25.04 | 0.53 | 1.00 |
| | Gmed | 94.92 | 95.81 | 2.01 | 104.32 | 105.99 | 2.83 | 0.99 | 29.22 | 29.56 | 0.85 | 34.75 | 35.08 | 0.84 | 1.00 |
| | Gmin | 21.72 | 21.98 | 1.70 | 22.52 | 22.32 | 1.53 | 1.69 | 32.74 | 32.86 | 1.67 | 38.11 | 38.05 | 1.54 | 0.99 |
| | Iliacus | 34.14 | 33.99 | 1.19 | 37.95 | 37.91 | 1.24 | 0.99 | 50.99 | 52.43 | 1.69 | 53.45 | 54.51 | 1.37 | 0.97 |
| | Obtext | 12.95 | 12.79 | 0.81 | 13.37 | 13.22 | 0.90 | 0.96 | 30.11 | 31.06 | 2.29 | 34.98 | 35.93 | 2.30 | 0.97 |
| | Obtint | 13.82 | 13.22 | 0.73 | 14.72 | 14.23 | 0.61 | 0.97 | 39.65 | 40.31 | 1.48 | 43.53 | 44.19 | 1.40 | 0.99 |
| | Pect | 11.07 | 10.95 | 0.58 | 11.47 | 11.34 | 0.48 | 0.98 | 36.86 | 38.03 | 1.64 | 39.57 | 40.37 | 1.37 | 0.97 |
| | Piri | 8.25 | 7.75 | 1.24 | 9.16 | 8.82 | 1.17 | 0.90 | 28.06 | 31.36 | 4.70 | 32.35 | 34.93 | 4.26 | 0.81 |
| | Pmajor | 19.35 | 18.56 | 1.43 | 21.96 | 20.93 | 1.67 | 0.97 | 42.08 | 43.46 | 1.55 | 44.03 | 45.36 | 1.50 | 0.98 |
| Thigh muscles | Adds | 188.02 | 189.11 | 3.82 | 210.85 | 211.59 | 4.31 | 1.00 | 36.63 | 36.82 | 0.61 | 39.67 | 39.81 | 0.49 | 1.00 |
| | Biceps | 71.30 | 71.31 | 1.85 | 76.28 | 76.60 | 2.64 | 0.99 | 34.23 | 34.40 | 0.78 | 36.26 | 36.39 | 0.80 | 1.00 |
| | Gracil | 19.66 | 19.70 | 0.77 | 20.44 | 20.37 | 0.96 | 0.98 | 27.37 | 28.40 | 1.42 | 28.52 | 29.56 | 1.42 | 0.99 |
| | Recfem | 49.35 | 49.10 | 1.50 | 54.33 | 53.94 | 1.59 | 0.99 | 45.66 | 46.15 | 0.57 | 45.62 | 46.01 | 0.51 | 0.99 |
| | Sart | 38.97 | 38.76 | 1.04 | 39.28 | 39.17 | 1.25 | 0.99 | 30.20 | 31.15 | 1.16 | 30.76 | 31.71 | 1.23 | 0.99 |
| | Semim | 50.26 | 50.63 | 2.39 | 54.53 | 54.90 | 2.71 | 0.97 | 29.69 | 30.06 | 0.86 | 33.50 | 33.74 | 0.70 | 1.00 |
| | Semit | 41.66 | 41.78 | 1.83 | 44.03 | 44.41 | 2.23 | 0.98 | 34.19 | 34.80 | 1.46 | 36.59 | 37.16 | 1.44 | 0.98 |
| | Tenfas | 19.80 | 19.85 | 0.95 | 19.89 | 19.94 | 0.97 | 0.98 | 23.85 | 24.83 | 1.31 | 25.86 | 27.01 | 1.40 | 0.99 |
| | Vaslatint | 237.39 | 238.48 | 4.50 | 253.34 | 253.95 | 3.82 | 1.00 | 47.84 | 48.12 | 0.65 | 49.93 | 50.17 | 0.59 | 1.00 |
| | Vasmed | 104.64 | 105.33 | 3.06 | 113.61 | 115.50 | 2.96 | 0.98 | 45.65 | 45.55 | 0.63 | 46.67 | 46.58 | 0.61 | 0.99 |
| Bones | Pelvis | 122.26 | 122.47 | 0.85 | 121.49 | 121.63 | 0.77 | 1.00 | 320.95 | 321.56 | 1.51 | 334.44 | 335.14 | 1.31 | 1.00 |
| | Femur | 169.62 | 170.27 | 1.27 | 169.65 | 170.23 | 1.07 | 1.00 | 447.29 | 447.35 | 1.51 | 464.81 | 464.79 | 1.40 | 1.00 |
| | Sacrum* | 94.22 | 94.00 | 1.74 | – | – | – | 0.97 | 190.39 | 192.60 | 3.34 | – | – | – | 0.99 |
| Mean±SD | | | | 1.77±1.06 | | | 1.89±1.17 | | | | 1.46±0.95 | | | 1.38±0.89 | |

* The measurements on the whole sacrum were reported since it was not separated into right/left.

**Table 7.** Impact of the number of training cases on the segmentation accuracy of the 6-layers model. DC: Dice coefficient, ASD: Average symmetric surface distance, n.s.: not significant, *: $p < .05$, **: $p < .01$.

| No. training cases | DC↑ | | ASD [mm]↓ | |
|---|---|---|---|---|
| 10 | .931±.034 | | 0.699±0.533 | |
| | | }** | | }* |
| 20 | .941±.019 | | 0.549±0.209 | |
| | | }** | | }* |
| 30 | .945±.014 | | 0.504±0.174 | |
| | | } n.s. | | } n.s. |
| 40 | .947±.013 | | 0.488±0.162 | |

**Table 8.** Impact of the number of dropout samples on the segmentation accuracy.

| No. dropout samples | DC↑ | ASD [mm]↓ |
|---|---|---|
| 1 | .941±.014 | 0.548±0.210 |
| 5 | .946±.013 | 0.526±0.189 |
| 10 | .947±.013 | 0.524±0.185 |
| 15 | .947±.013 | 0.525±0.186 |
| 20 | .947±.013 | 0.521±0.183 |
| 50 | .947±.013 | 0.524±0.185 |

Predictive uncertainty was investigated in several studies to predict the segmentation accuracy in medical images.[7,29–32]. Nowak et al. investigated the predictive uncertainty (entropy) in segmenting skeletal muscles in lumbar-level CT slices from dual centers with CT scanners from multiple manufacturers. Their study showed the applicability of the predictive uncertainty on the data from both centers; however, it was only applied to 2D CT slices, with the muscles combined into a single label. Mehtrash et al. investigated the predictive uncertainty (entropy) based on ensemble models. The method was validated on multiple structures at MRIs, and strong correlations between the predictive uncertainty and segmentation accuracy were reported. However, both studies did not address the segmentation of individual muscles or bones and did not investigate the impact of practically important factors, such as disease condition or numbers of training data on the segmentation accuracy[29,30].

In our experiments, we attempted to use the entropy of single samples and observed slightly improved correlations with the segmentation accuracy. However, the segmentation accuracy has decreased. Indeed, larger numbers of 10 samples seem to improve the overall accuracy (See 8). Compared with the ensemble approach[29], the MCDropout approach showed a good balance between the segmentation accuracy, computation time, and accuracy of the predictive uncertainty.

Compared with the baseline model[7], this study showed a potential improvement when increasing the depth of the segmentation model and the number of training data. Increasing the training data to larger than 20 cases improved overall, as shown in Fig. 4. Other studies have also investigated the segmentation of thigh muscles from CT images[45–47]. However, the number of cases was smaller, making the comparison invalid. Recently, Kim et al. attempted a 3D UNETR[44] for the segmentation of the full thigh muscles[19]. The model was trained on a larger dataset (60 cases) and tested on 12 cases; however, the dataset included only patients with hip fractures, and it showed lower accuracy (DC=0.84; ASSD=1.419±0.91 mm). These comparisons collectively emphasize the higher accuracy of the improved model and the uniqueness of our fully annotated database (DB#1) and validation of external databases (DB#2-5) regarding the number of cases and the diversity of disease, patient positioning, and imaging conditions.

The assessment of the volume and intensity of the muscles and bones are among the ultimate goals of automated MSK image segmentation. In particular, the mean HU measured at abdominal muscles has shown a higher potential to predict age-related adverse outcomes compared with the muscle area[9]. To our knowledge, this is the first study to investigate the accuracy of these measurements in automatically segmented hip and thigh MSK structures in CT images. High accuracy of the volumes and mean HU of most muscles and bones in HOA patients was obtained. Furthermore, the validation experiment on the four databases showed the robustness of the improved model in the segmentation of Gmed muscle with respect to the multi-manufacturer/scanners and disease variations. These findings indicate the potential usability of the segmentation model for hip-to-knee MSK assessments in clinical routines. The rapid inference time (~3 min) of the entire CT volume adds to the model's practicality for adoption in surgical planning or musculoskeletal simulation platforms. Furthermore, the musclewise density visualization depicted in Fig. 7 would help in the rapid and comprehensive assessment of muscle quality under several conditions, such as HOA, cancer, sarcopenia, and obesity[20].

On the other hand, MSK segmentation approaches in magnetic resonance images (MRIs) are attracting attention due to patient safety and high soft tissue contrast[48,49], and the possibility of quantifying the muscle/fat composition using special sequences, such as Dixon[50]. However, MRIs usually require a long scanning time, represent various characteristics based on the acquisition sequence, and cover limited fields of view (FOVs). This necessitates integrating multiple acquisitions and registration processing to assess the whole knee-to-hip[49], which could be limited to a few research-purposed databases[22].

This study has the following limitations. The 2D segmentation model, even though it has a rapid inference time, does not capture the 3D information of neighboring structures, which affects the segmentation of small structures, such as the Piri muscle. Furthermore, the small hip muscles (ObtInt and ObtExt) showed low AUROC in failure detection based on the predictive uncertainty. The usage of the auto segmentations of those muscles requires attention in *unannotated* databases.

The failure detection approach and improved model (*6layers,50*) create a basis for several future directions in our research. The model's potential in analyzing the disease progression of individual bones and muscles in large-scale databases of *unannotated* CTs will be investigated. Cases with segmentation failures could be detected based on the predictive uncertainty and excluded or refined by human annotators for downstream MSK analyses. Furthermore, the extension of the segmentation model to predict the MSK structures in other regions, such as the abdomen and back muscles, is currently under development. Furthermore, a few muscles in the hip, such as quadratus femoris and Gemelli muscles, were not addressed, besides combining several muscles, such as the adductors, into a single label due to the challenging boundary definition. These structures will be addressed in our future work by involving higher-resolution images, such as from photon-counting CTs.

## 5 Conclusions

This study validated a DL model for MSK segmentation with uncertainty estimation in clinical CT images. The improved model (*6layers,50*) allowed for the automated, rapid, and accurate assessment of the volume and density of the hip and thigh bones and muscles from clinical CT images. It has shown the usability of the predictive uncertainty as a tool for predicting the segmentation accuracy and failure detection in individual MSK structures at *unannotated* CT image databases.The high segmentation and muscle volume/density estimation accuracy, along with the high accuracy in failure detection, exhibited the model's reliability for the analysis of individual MSK structures in large-scale CT databases.

## 6 Data Availability

The datasets used and analyzed in the current study are available from the corresponding author upon reasonable request.

## Acknowledgements

This study was supported by Japan Society for the Promotion of Science (JSPS) KAKENHI (grant numbers JP19H01176, JP20H04550, JP21K16655, JP21K18080, JP17H04266, JP17K16482, JP20K08056, JP21H03799, and JP23K07214), Uehara Memorial Foundation, and the Takeda Science Foundation.

## Author contributions statement

M.S. drafted the manuscript, conducted the experiments, and performed the data analysis. Y.O., M.I., K.U., M.T., N.S, and Y.S. made a substantial contribution to the study design and data collection. T.H., M.H., Y.Y., M.Y., Y.Y., S.O., M.J. contributed to data collection. All authors reviewed the manuscript.

## Additional information

## Competing interests

Masahiro Jinzaki received a grant from Canon Medical Systems. However, Canon Medical Systems was not involved in the design and conduct of the study; in the collection, analysis, and interpretation of the data; or in the preparation, review, and approval of the manuscript. The remaining authors have no conflicts of interest to declare.

## Declarations

**Ethics approval** Ethical approval was obtained from the Institutional Review Boards (IRBs) of the institutions participating in this study (IRB approval numbers: 21115 for Osaka University Hospital, 2023-28 for Hitachi Health Care Center, 2020-M-7 for Nara Institute of Science and Technology, and jRCTs032180267 for Keio University.)

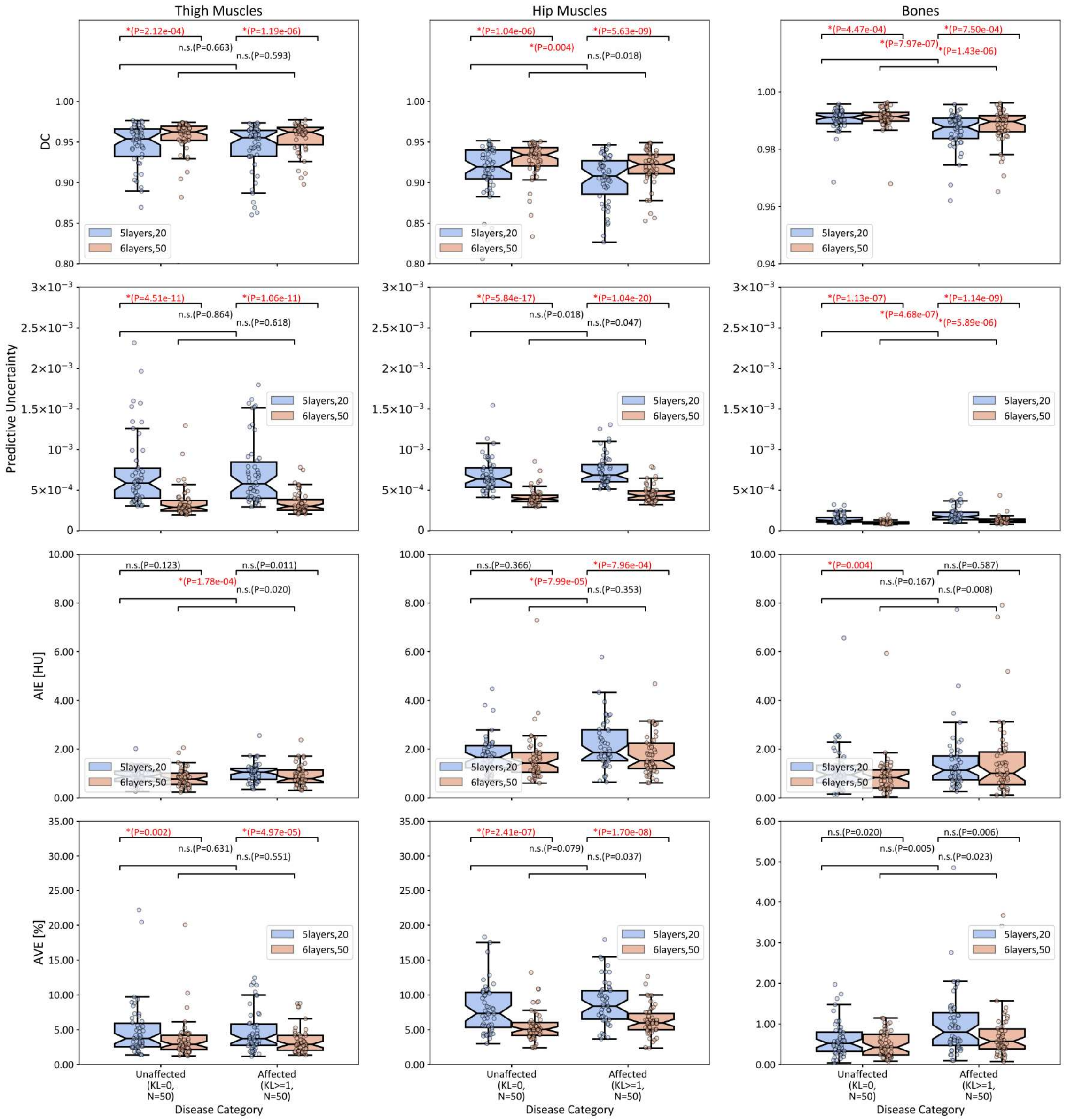

(a) Internal training/validation database.

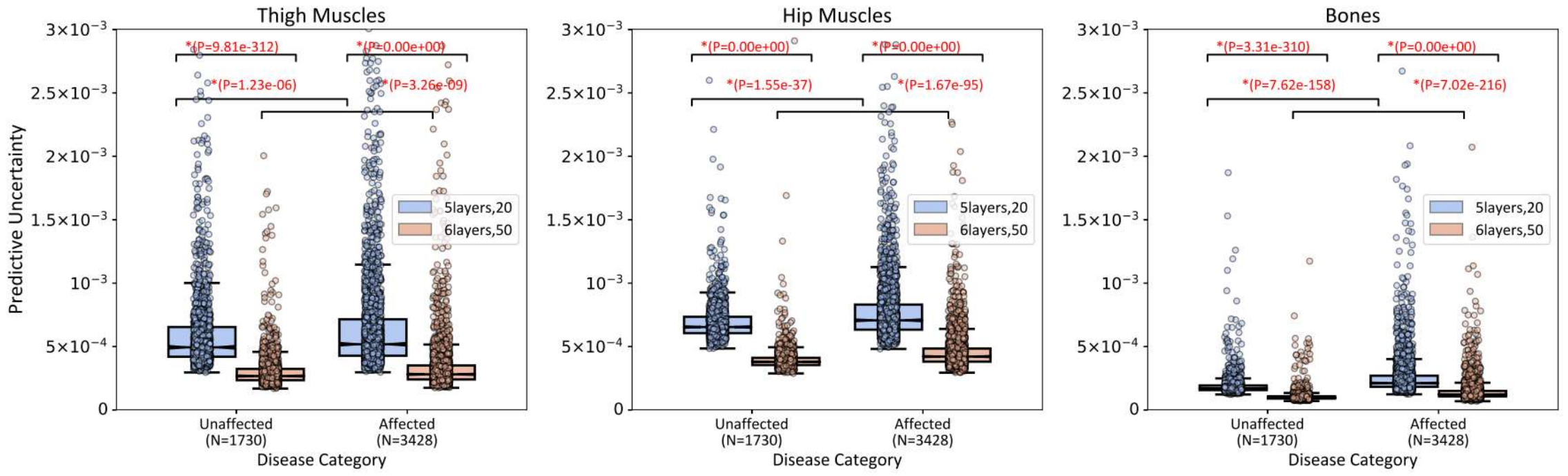

(b) External large-scale validation database.

**Figure 5.** Distributions of the accuracy evaluation metrics and predictive uncertainty of the three MSK structure groups, i.e., thigh (left) and hip (middle) muscles and bones (right), in terms of the disease status of body sides in HOA patients in internal validation DB#1 (a) and external validation large-scale DB#5 (b). N: number of cases. n.s.: not significant, *: $p < .004$. (Based on Shapiro's normality test, the hypothesis test was performed using either the Wilcoxon signed-rank test or the Student's t-test. Bonferroni correction was used for the multiple comparisons.)

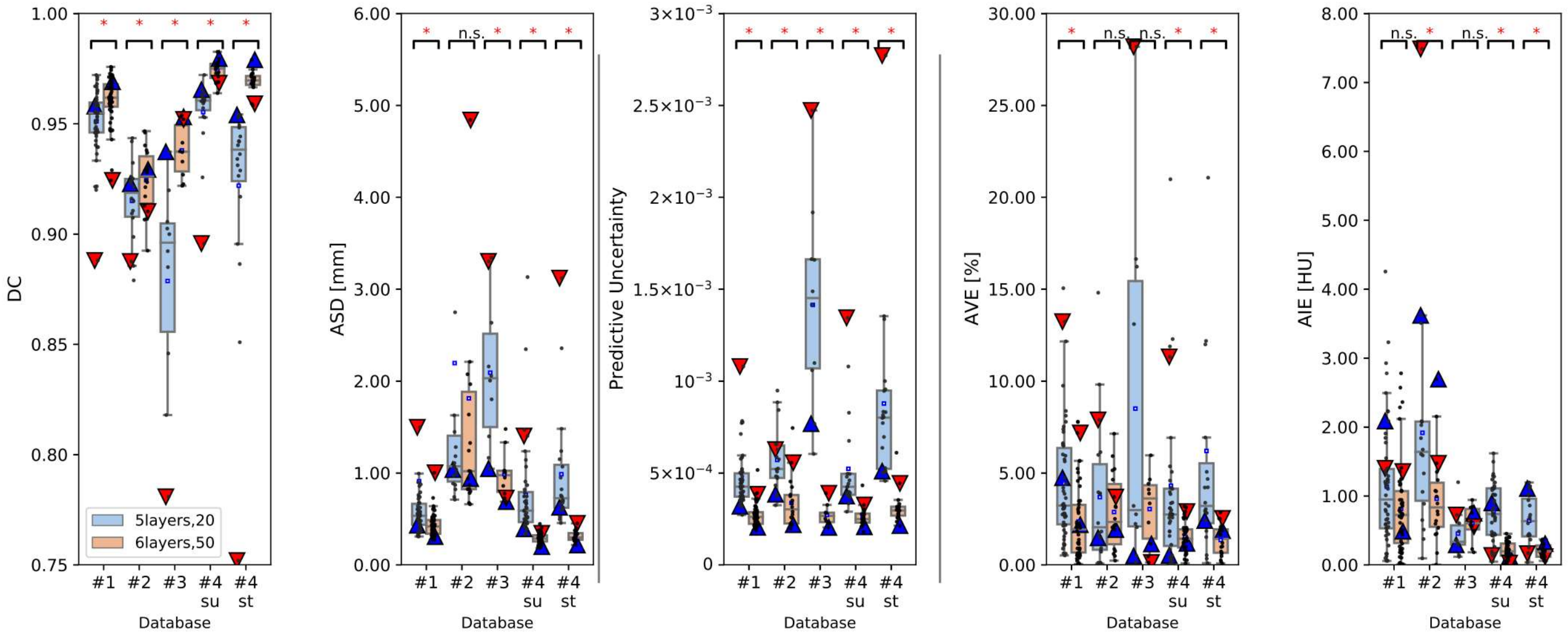

(a) Segmentation accuracy (b) Predictive uncertainty (c) Volume/mean HU accuracy

**Figure 6.** Comparison between segmentation model accuracy (a, c) and predictive uncertainty (b) of Gmed in the multi-manufacturer/scanner databases. DC: Dice coefficient, ASD: Average symmetric surface distance, AVE: Average volume error, AIE: Average intensity error, n.s.: not significant, **: $p < .01$,***: $p < .001$. (Based on Shapiro's normality test, the hypothesis tests were performed using either the Wilcoxon signed-rank test or the Student's t-test). The triangles indicate the cases corresponding to the 5th (▲) and 95th (▼) quantiles of the predictive uncertainty visualized in A.7 and A.8.

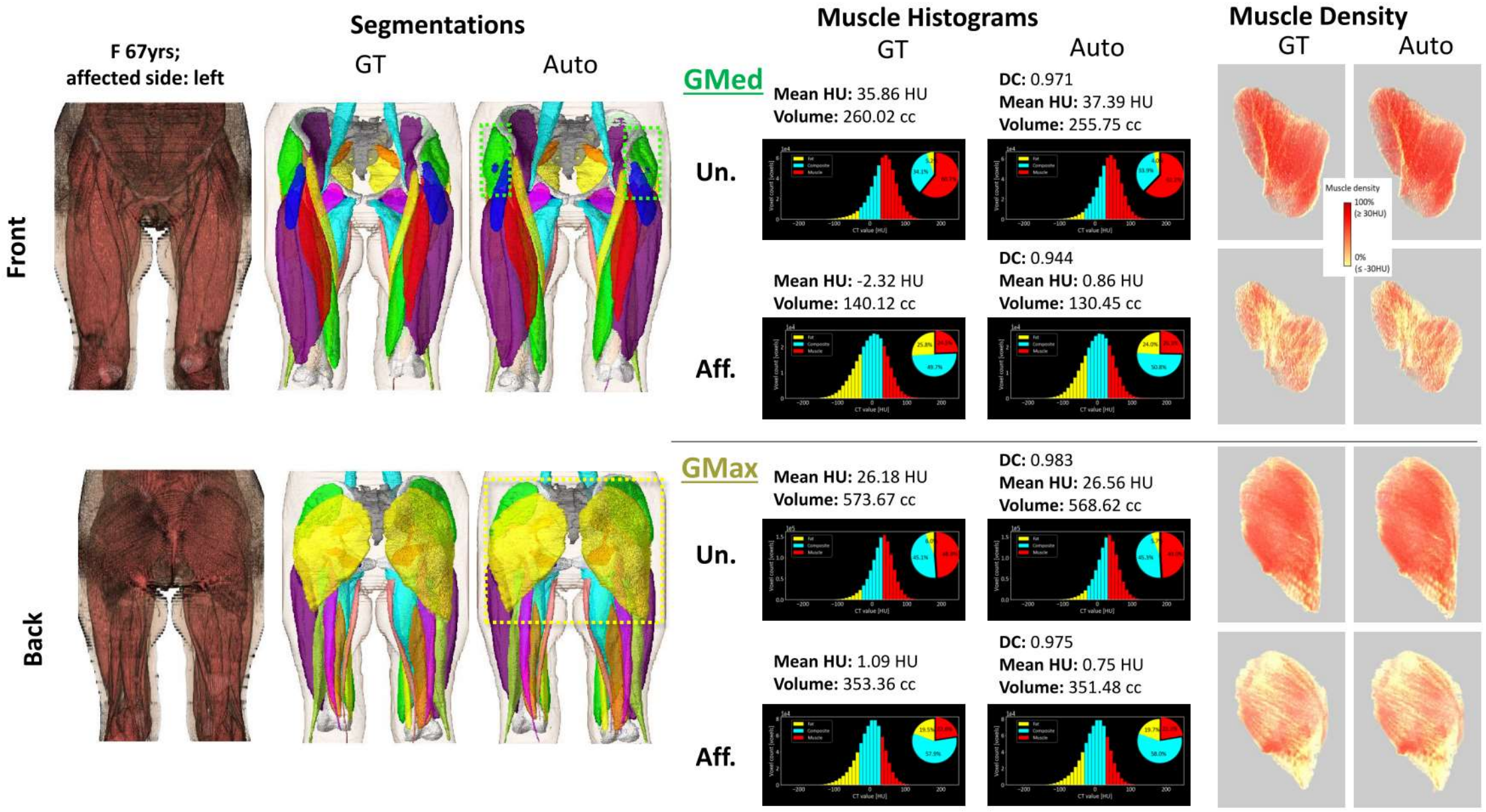

**Figure 7.** Ground-truth (GT) and predicted (Auto) segmentations of the unaffected (Un.) and affected (Aff.) sides of a representative HOA case (median DC in Fig.3) with diagnostic biomarkers, histograms, and muscle density visualizations of the Gmax and Gmed muscles.

# Appendices

## A Muscle-wise evaluation metrics

Figures A.1-A.5 summarize the evaluation metrics and predictive uncertainty of each hip-to-knee structure (3 bones and 19 muscles) using the three models from the 5-fold cross-validation experiments. The box plots include the detailed metric values of the box plots depicted in Fig.3.

- Fig.A.1: Dice coefficient (DC), corresponding to Fig.3(a).
- Fig.A.2: Average symmetric surface distance (ASD, mm), corresponding to Fig.3(a).
- Fig.A.3: Predictive uncertainty, corresponding to Fig.3(b).
- Fig.A.4: Average volume error (AVE,%), corresponding to Fig.3(c).
- Fig.A.5: Average intensity error (AIE,HU), corresponding to Fig.3(c).

## B Relationship between predictive uncertainty and segmentation accuracy (Dice coefficient).

Figure A.6 depicts the relationship between the predictive uncertainty and segmentation accuracy (DC) in terms of a) the segmentation model and b) the number of training images. The two plots correspond to the box plots in Fig. 3 and Table 7, respectively. Each point represents a single muscle/bone (right and left sides combined). All experiments were performed at DB#1.

## C Representative segmentation results from the four databases used in the study

Figure A.7 and A.8 shows segmentation results from the databases DB#1-4 used in this study, with qualitative and quantitative evaluations of the Gmed and other structures. In each case, the models *5layers,20* and *6layers,50* predicted the upper and lower segmentations, respectively. For each database, the upper and lower cases correspond with the $5^{th}$ (▲) and $95^{th}$ (▼) quantiles of the predictive uncertainty visualized in Fig. 6. The cases with lower uncertainty had higher segmentation accuracy in both models and vice versa.

## D Usability of the relationship between predictive uncertainty and segmentation accuracy for detection of inaccurate and failed segmentations.

Figures A.9 and A.10 show the relationships between the predictive uncertainty and segmentation accuracy for detecting inaccurate and failed segmentations.

- Fig.A.9: Each structure in DB#1.
- Fig.A.10: Gmed in the four databases DB#1-4.

## E Usability of the predictive uncertainty for detecting inaccurate and failed segmentations in large-scale databases.

Figure A.11 shows the usability of the predictive uncertainty in detecting inaccurate and failed segmentations in a large-scale database consisting of 2579 CT images. The scatter plot shows the average predictive uncertainty of all structures by the *5layers,20* and *6layers,50* models. The right side depicts representative cases selected based on the thresholds shown in 4. Most cases have shown lower predictive uncertainty by the *6layers,50* model. The detected segmentation failure case (⋆) had a large variation in the hip positioning, possibly caused by the patient's discomfort due to the disease.

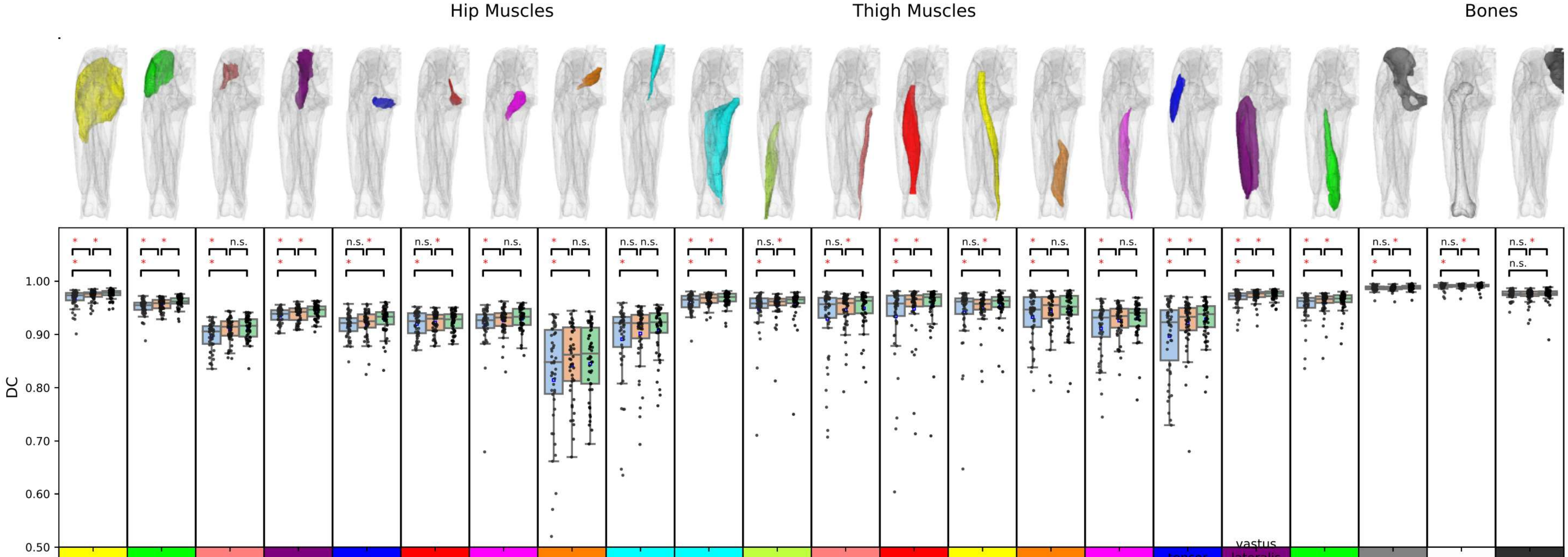

| | gluteus maximus | gluteus medius | gluteus minimus | iliacus | obturator externus | obturator internus | pectineus | piriformis | psoas major | adductors | biceps femoris | gracilis | rectus femoris | sartorius | semi membranosus | semi tendinosus | tensor fasciae latae | vastus lateralis and vastus | vastus medialis | pelvis | femur | sacrum |
|---|---|---|---|---|---|---|---|---|---|---|---|---|---|---|---|---|---|---|---|---|---|---|
| 5layers,20 (conv.) | 0.968 ±0.016 | 0.951 ±0.015 | 0.897 ±0.026 | 0.935 ±0.014 | 0.917 ±0.022 | 0.920 ±0.023 | 0.918 ±0.041 | 0.814 ±0.144 | 0.891 ±0.071 | 0.960 ±0.017 | 0.948 ±0.042 | 0.929 ±0.067 | 0.934 ±0.072 | 0.941 ±0.055 | 0.933 ±0.046 | 0.911 ±0.051 | 0.897 ±0.068 | 0.968 ±0.016 | 0.954 ±0.028 | 0.987 ±0.005 | 0.990 ±0.005 | **0.975 ±0.009** |
| 5layers,50 | 0.973 ±0.010 | 0.956 ±0.011 | 0.909 ±0.023 | 0.938 ±0.015 | 0.920 ±0.026 | 0.922 ±0.020 | 0.926 ±0.025 | 0.842 ±0.100 | 0.902 ±0.057 | 0.965 ±0.011 | 0.955 ±0.026 | 0.946 ±0.037 | 0.948 ±0.052 | 0.950 ±0.032 | **0.944 ±0.037** | 0.926 ±0.031 | 0.921 ±0.048 | 0.973 ±0.012 | 0.960 ±0.022 | 0.986 ±0.005 | **0.991 ±0.004** | 0.973 ±0.013 |
| 6layers,50 | **0.975 ±0.009** | **0.961 ±0.011** | **0.912 ±0.021** | **0.943 ±0.014** | **0.927 ±0.025** | **0.925 ±0.019** | **0.929 ±0.022** | **0.845 ±0.091** | **0.908 ±0.046** | **0.967 ±0.013** | **0.957 ±0.033** | **0.949 ±0.036** | **0.954 ±0.045** | **0.955 ±0.029** | **0.944 ±0.040** | **0.930 ±0.034** | **0.929 ±0.036** | **0.975 ±0.009** | **0.963 ±0.018** | **0.988 ±0.005** | **0.991 ±0.005** | **0.975 ±0.014** |

**Figure A.1.** Dice Coefficient (DC), corresponds to Fig. 3(a).

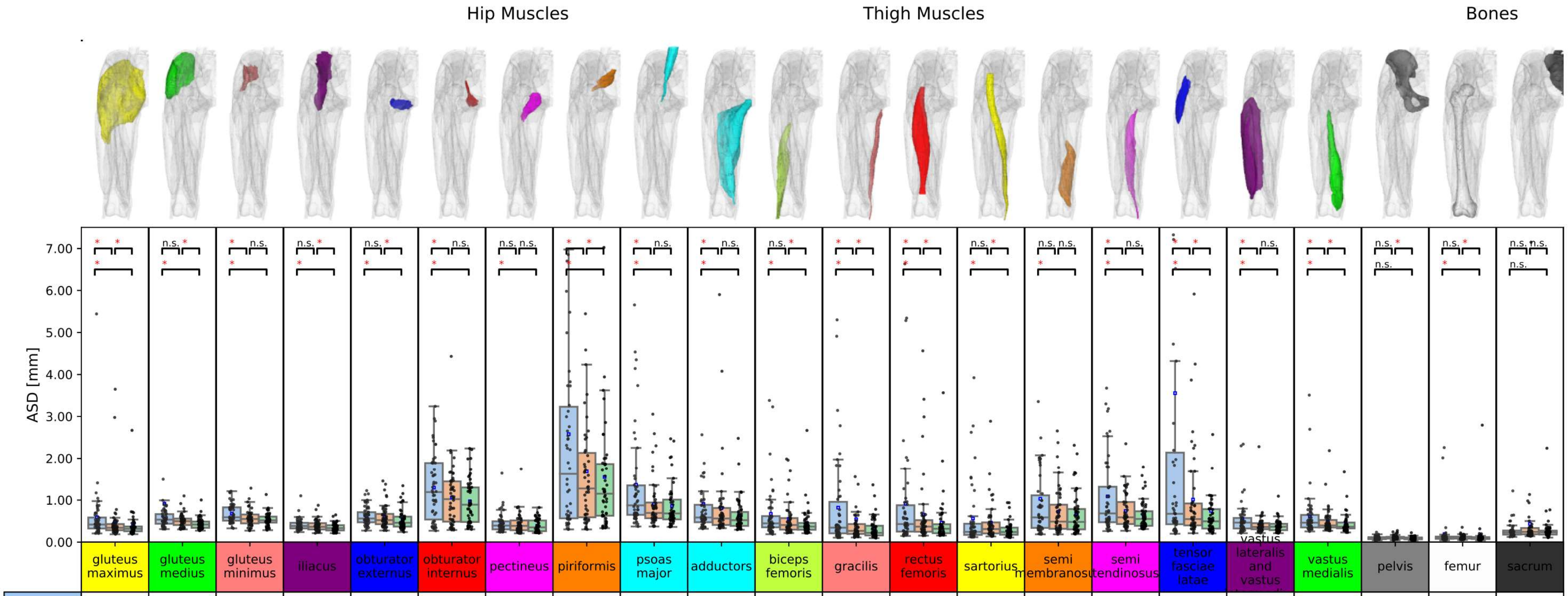

| | gluteus maximus | gluteus medius | gluteus minimus | iliacus | obturator externus | obturator internus | pectineus | piriformis | psoas major | adductors | biceps femoris | gracilis | rectus femoris | sartorius | semi membranosus | semi tendinosus | tensor fasciae latae | vastus lateralis and vastus intermedius | vastus medialis | pelvis | femur | sacrum |
|---|---|---|---|---|---|---|---|---|---|---|---|---|---|---|---|---|---|---|---|---|---|---|
| 5layers,20 (conv.) | 0.588 ±0.748 | 0.910 ±2.436 | 0.676 ±0.221 | 0.412 ±0.141 | 0.623 ±0.223 | 1.296 ±0.804 | 0.454 ±0.243 | 2.577 ±3.006 | 1.364 ±1.207 | 0.907 ±1.165 | 0.670 ±0.654 | 0.818 ±1.131 | 0.914 ±1.348 | 0.560 ±0.785 | 1.029 ±1.374 | 1.086 ±0.918 | 3.554 ±6.729 | 0.558 ±0.424 | 0.610 ±0.565 | 0.097 ±0.039 | 0.190 ±0.405 | **0.267 ±0.178** |
| 5layers,50 | 0.482 ±0.600 | 0.505 ±0.158 | 0.569 ±0.194 | 0.392 ±0.130 | 0.599 ±0.267 | 1.062 ±0.728 | 0.437 ±0.252 | 1.684 ±1.580 | 0.903 ±0.580 | 0.807 ±0.948 | 0.549 ±0.416 | 0.542 ±1.370 | 0.651 ±0.880 | 0.460 ±0.472 | 0.723 ±0.573 | 0.739 ±0.423 | 1.011 ±1.137 | 0.426 ±0.292 | 0.490 ±0.313 | 0.113 ±0.053 | **0.113 ±0.052** | 0.431 ±0.991 |
| 6layers,50 | **0.390 ±0.354** | **0.432 ±0.132** | **0.537 ±0.154** | **0.348 ±0.096** | **0.532 ±0.241** | **0.970 ±0.535** | **0.389 ±0.156** | **1.549 ±1.500** | **0.861 ±0.518** | **0.633 ±0.399** | **0.451 ±0.368** | **0.322 ±0.254** | **0.462 ±0.547** | **0.290 ±0.188** | **0.659 ±0.502** | **0.642 ±0.343** | **0.743 ±1.125** | **0.382 ±0.150** | **0.431 ±0.237** | **0.091 ±0.037** | 0.152 ±0.384 | 0.292 ±0.309 |

**Figure A.2.** Average symmetric surface distance (ASD), corresponds to Fig. 3(a).

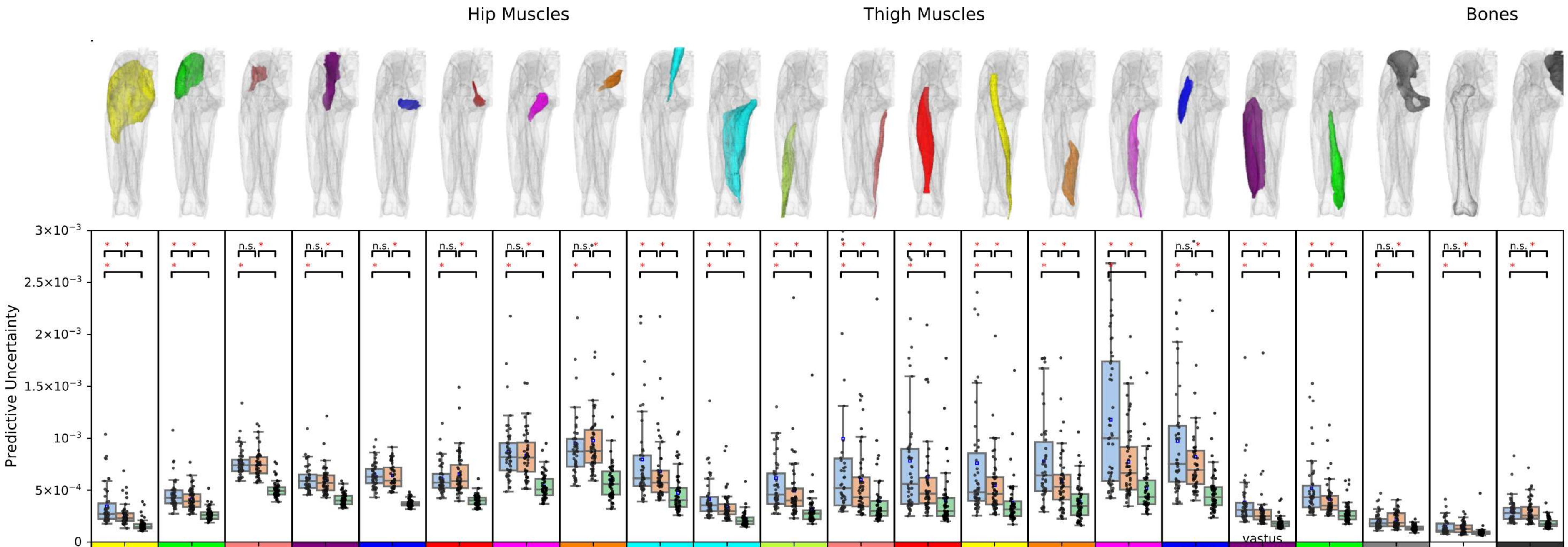

| | gluteus maximus | gluteus medius | gluteus minimus | iliacus | obturator externus | obturator internus | pectineus | piriformis | psoas major | adductors | biceps femoris | gracilis | rectus femoris | sartorius | semi membranosus | semi tendinosus | tensor fasciae latae | vastus lateralis and vastus | vastus medialis | pelvis | femur | sacrum |
|---|---|---|---|---|---|---|---|---|---|---|---|---|---|---|---|---|---|---|---|---|---|---|
| 5layers,20 (conv.) | 3.4e-04 ±1.9e-04 | 4.6e-04 ±1.5e-04 | 7.6e-04 ±1.4e-04 | 6.1e-04 ±1.3e-04 | 6.4e-04 ±1.2e-04 | 6.1e-04 ±1.4e-04 | 8.7e-04 ±3.0e-04 | 9.5e-04 ±4.3e-04 | 8.0e-04 ±4.5e-04 | 4.1e-04 ±2.1e-04 | 6.1e-04 ±5.8e-04 | 9.9e-04 ±1.1e-03 | 7.8e-04 ±6.0e-04 | 7.6e-04 ±6.7e-04 | 7.8e-04 ±4.4e-04 | 1.2e-03 ±6.7e-04 | 9.7e-04 ±5.7e-04 | 3.8e-04 ±2.6e-04 | 5.1e-04 ±2.9e-04 | 2.0e-04 ±7.6e-05 | 1.5e-04 ±7.3e-05 | 2.9e-04 ±1.2e-04 |
| 5layers,50 | 2.6e-04 ±1.0e-04 | 4.1e-04 ±9.9e-05 | 7.5e-04 ±1.3e-04 | 5.9e-04 ±1.4e-04 | 6.4e-04 ±1.2e-04 | 6.5e-04 ±2.1e-04 | 8.3e-04 ±2.3e-04 | 9.8e-04 ±3.8e-04 | 6.8e-04 ±3.3e-04 | 3.4e-04 ±1.4e-04 | 4.9e-04 ±3.3e-04 | 6.0e-04 ±4.7e-04 | 6.3e-04 ±4.8e-04 | 5.5e-04 ±3.0e-04 | 5.8e-04 ±3.0e-04 | 7.7e-04 ±3.6e-04 | 8.2e-04 ±4.7e-04 | 3.1e-04 ±2.4e-04 | 4.1e-04 ±1.7e-04 | 2.2e-04 ±7.7e-05 | 1.3e-04 ±5.2e-05 | 3.0e-04 ±1.1e-04 |
| 6layers,50 | **1.6e-04 ±5.6e-05** | **2.7e-04 ±6.0e-05** | **5.1e-04 ±8.0e-05** | **4.1e-04 ±6.0e-05** | **3.7e-04 ±3.6e-05** | **4.0e-04 ±5.4e-05** | **5.3e-04 ±1.3e-04** | **6.0e-04 ±2.2e-04** | **4.7e-04 ±2.0e-04** | **2.2e-04 ±7.2e-05** | **3.1e-04 ±2.1e-04** | **3.9e-04 ±3.3e-04** | **4.0e-04 ±2.9e-04** | **3.8e-04 ±2.4e-04** | **3.9e-04 ±1.9e-04** | **5.2e-04 ±2.5e-04** | **5.1e-04 ±3.1e-04** | **1.9e-04 ±5.6e-05** | **2.8e-04 ±9.8e-05** | **1.3e-04 ±2.6e-05** | **9.7e-05 ±5.6e-05** | **1.8e-04 ±4.9e-05** |

**Figure A.3.** Predictive uncertainty, corresponds to Fig. 3(b).

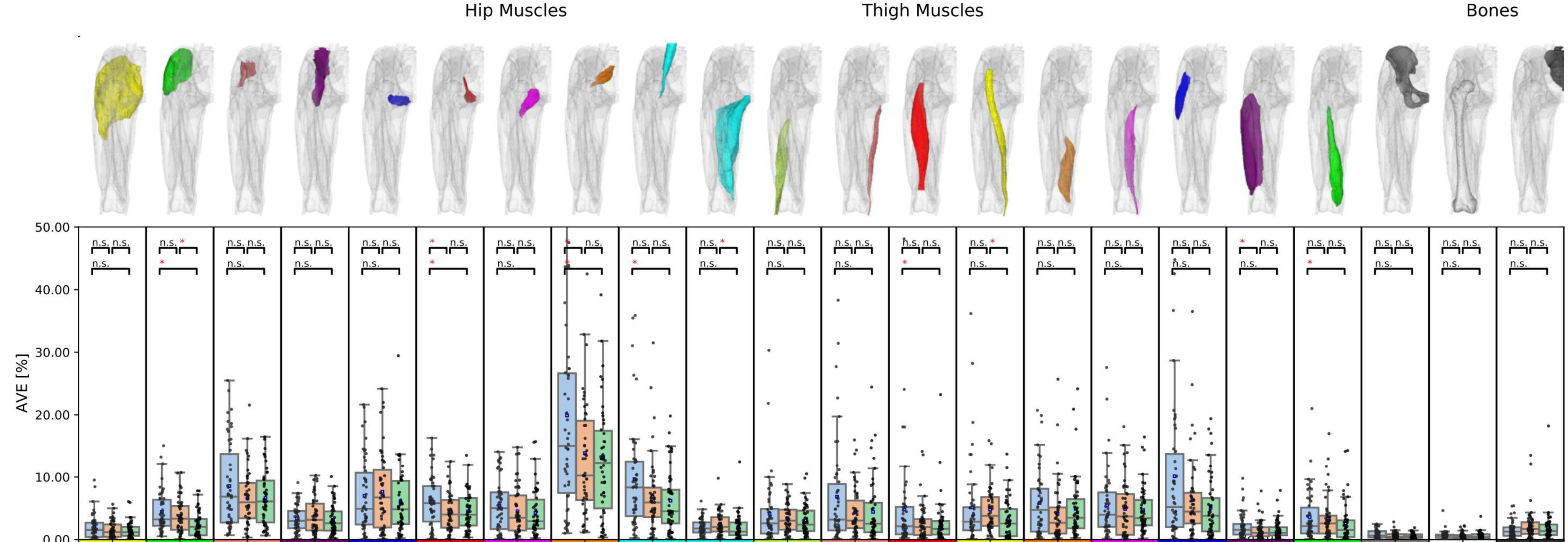

| | gluteus maximus | gluteus medius | gluteus minimus | iliacus | obturator externus | obturator internus | pectineus | piriformis | psoas major | adductors | biceps femoris | gracilis | rectus femoris | sartorius | semi membranosus | semi tendinosus | tensor fasciae latae | lateralis and vastus | vastus medialis | pelvis | femur | sacrum |
|---|---|---|---|---|---|---|---|---|---|---|---|---|---|---|---|---|---|---|---|---|---|---|
| 5layers,20 (conv.) | 1.994 ±1.988 | 4.426 ±3.303 | 8.532 ±6.854 | 3.403 ±2.168 | 6.970 ±5.963 | 5.963 ±4.105 | 5.395 ±4.278 | 19.884 ±17.942 | 9.653 ±8.495 | 2.021 ±1.350 | 4.156 ±5.376 | 6.842 ±8.389 | 4.778 ±7.958 | 5.145 ±6.968 | 5.831 ±5.336 | 5.473 ±5.435 | 10.121 ±12.159 | 2.145 ±2.063 | 3.572 ±3.956 | 0.830 ±0.692 | 0.669 ±0.693 | **1.493 ±1.248** |
| 5layers,50 | 1.520 ±1.352 | 3.727 ±2.673 | **6.575 ±4.480** | 3.791 ±2.849 | 7.621 ±6.196 | **4.441 ±3.097** | 4.511 ±3.826 | 13.883 ±12.800 | 6.831 ±5.994 | 2.512 ±1.855 | 3.401 ±2.244 | **4.362 ±3.812** | 2.788 ±2.933 | 4.791 ±3.600 | **4.109 ±4.814** | 4.868 ±4.466 | 6.315 ±6.727 | **1.532 ±1.538** | 3.372 ±3.624 | **0.592 ±0.551** | **0.606 ±0.534** | 2.274 ±2.530 |
| 6layers,50 | **1.483 ±1.304** | **2.392 ±2.037** | 6.644 ±4.298 | **3.139 ±2.342** | **6.022 ±5.228** | 4.501 ±3.040 | **4.267 ±3.794** | **13.222 ±11.905** | **6.195 ±4.896** | **1.903 ±2.014** | **3.064 ±2.409** | 4.426 ±5.121 | **2.622 ±3.677** | **3.236 ±3.156** | 4.886 ±5.005 | **4.706 ±3.664** | **5.089 ±4.973** | 1.551 ±1.544 | **2.593 ±3.257** | 0.605 ±0.438 | **0.606 ±0.603** | 1.913 ±2.587 |

**Figure A.4.** Average volume error (AVE), corresponds to Fig. 3(c).

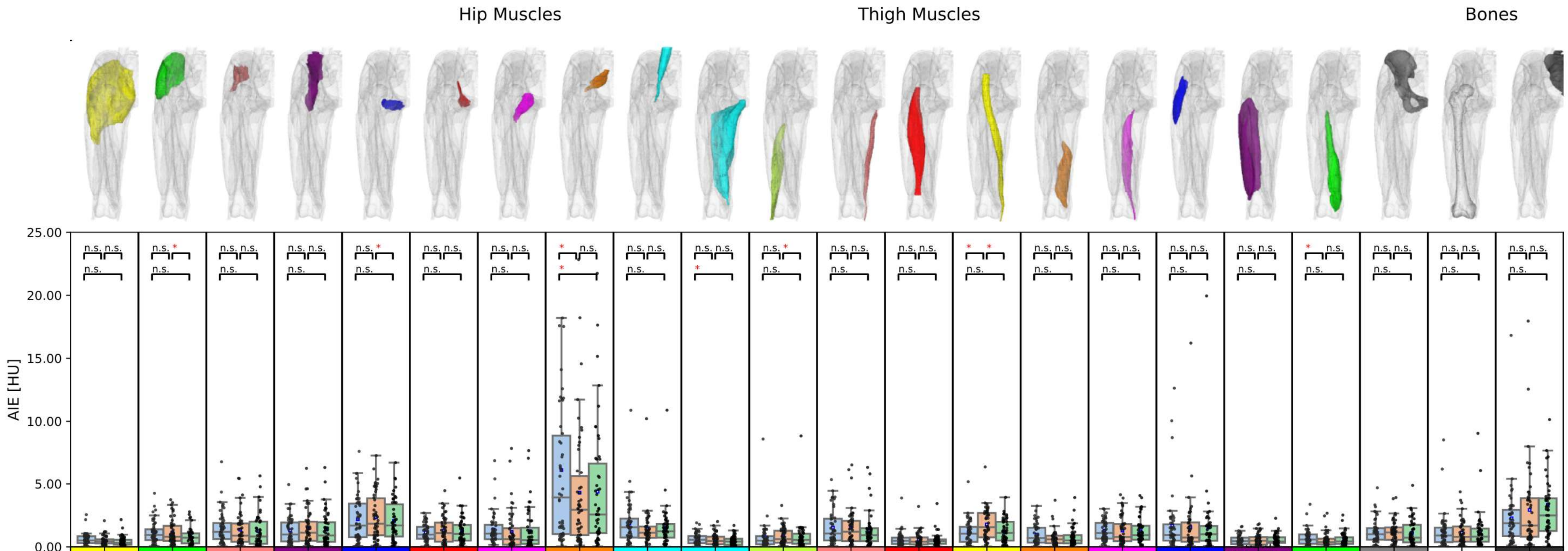

| | gluteus maximus | gluteus medius | gluteus minimus | iliacus | obturator externus | obturator internus | pectineus | piriformis | psoas major | adductors | biceps femoris | gracilis | rectus femoris | sartorius | semi membranosus | semi tendinosus | tensor fasciae latae | lateralis and vastus | vastus medialis | pelvis | femur | sacrum |
|---|---|---|---|---|---|---|---|---|---|---|---|---|---|---|---|---|---|---|---|---|---|---|
| 5layers,20 (conv.) | 0.590 ±0.483 | 1.132 ±0.906 | 1.490 ±1.259 | **1.306 ±1.078** | 2.196 ±1.792 | **1.137 ±0.685** | 1.371 ±1.274 | 6.130 ±7.426 | 1.854 ±1.813 | 0.635 ±0.457 | **0.754 ±1.240** | 1.513 ±1.357 | 0.589 ±0.602 | **1.139 ±0.950** | 0.970 ±0.815 | 1.279 ±0.799 | 1.674 ±2.426 | **0.524 ±0.406** | 0.718 ±0.644 | 1.199 ±0.990 | 1.226 ±1.446 | **2.269 ±2.506** |
| 5layers,50 | 0.494 ±0.381 | 1.152 ±0.983 | 1.360 ±1.297 | 1.481 ±1.348 | 2.359 ±1.852 | 1.294 ±1.061 | **1.190 ±1.651** | **4.300 ±4.790** | **1.398 ±1.457** | 0.572 ±0.453 | 0.862 ±0.681 | 1.506 ±1.451 | 0.568 ±0.556 | 1.774 ±1.209 | **0.695 ±0.655** | 1.295 ±1.106 | 1.559 ±2.407 | 0.573 ±0.462 | **0.516 ±0.554** | 1.212 ±0.871 | **1.114 ±1.001** | 2.928 ±3.274 |
| 6layers,50 | **0.417 ±0.393** | **0.803 ±0.669** | **1.240 ±1.276** | 1.381 ±1.352 | **2.056 ±1.632** | 1.243 ±1.042 | 1.251 ±1.675 | 4.376 ±4.806 | 1.509 ±1.637 | **0.470 ±0.373** | 0.794 ±1.255 | **1.270 ±1.359** | **0.517 ±0.541** | 1.206 ±0.923 | 0.763 ±0.751 | **1.258 ±1.012** | **1.529 ±2.812** | 0.565 ±0.415 | 0.558 ±0.514 | **1.107 ±1.002** | 1.190 ±1.496 | 3.337 ±4.124 |

**Figure A.5.** Average intensity error (AIE), corresponds to Fig. 3(c).

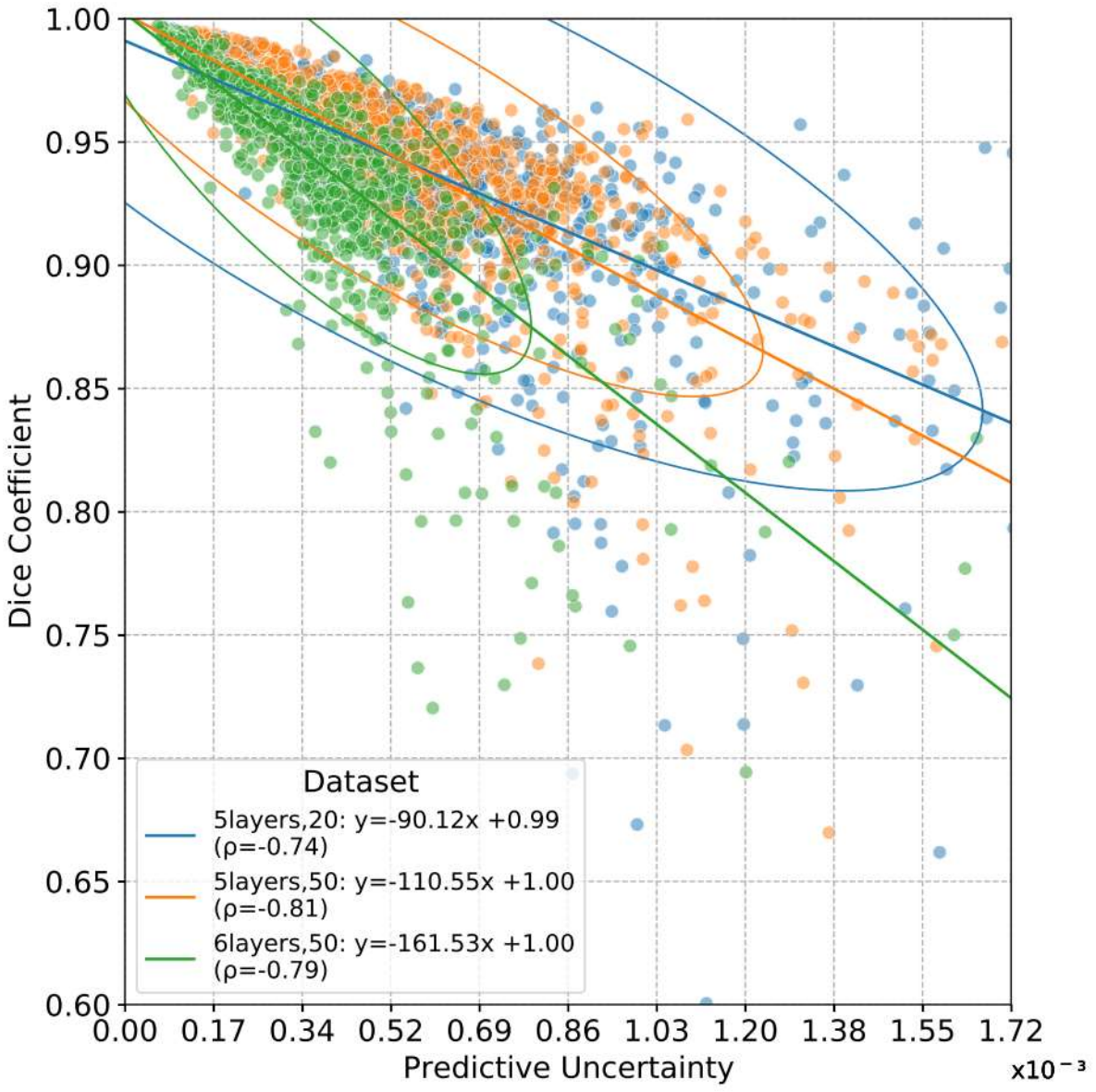

(a) Impact of segmentation models

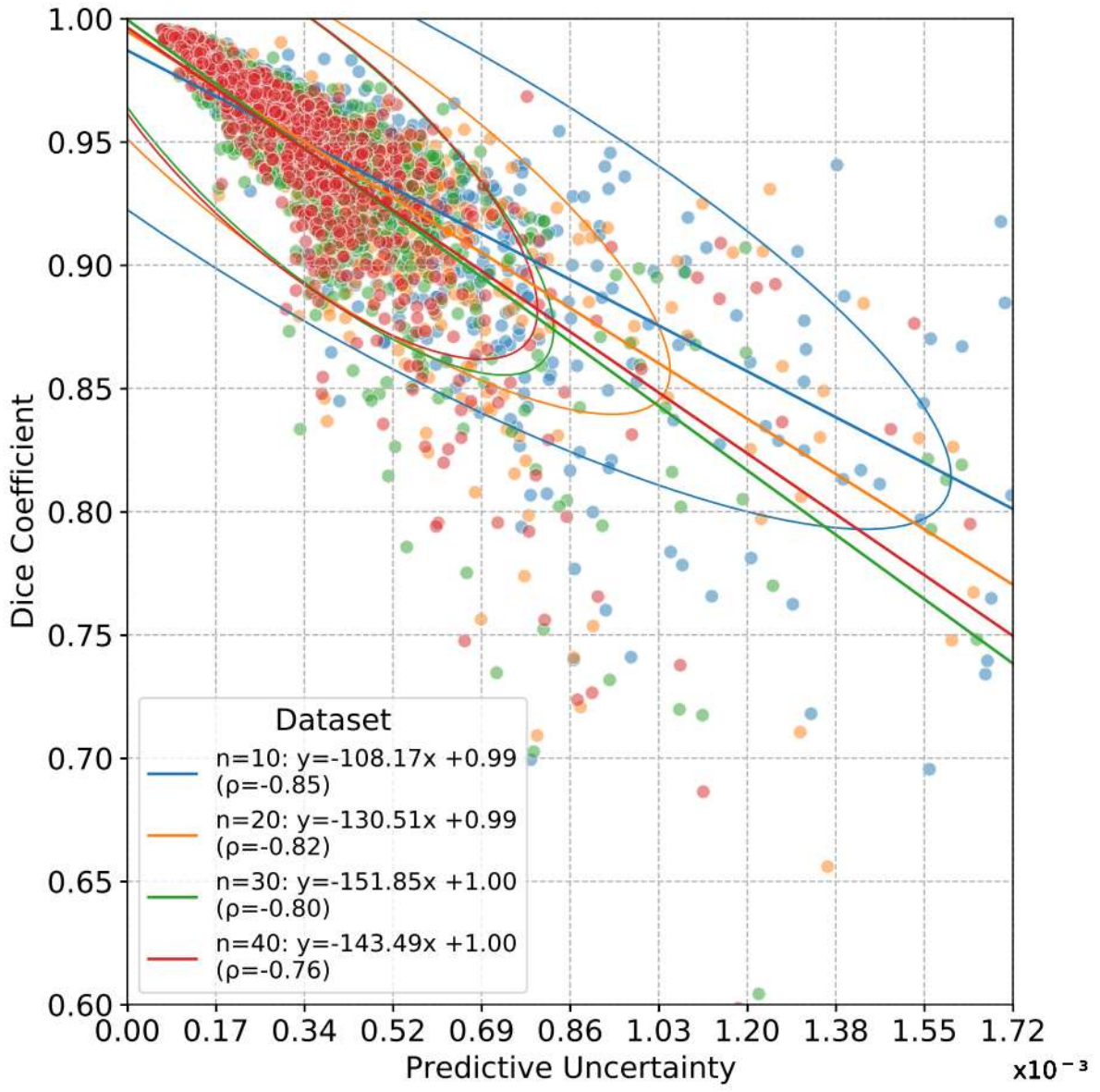

(b) Impact of number of training cases

**Figure A.6.** Relationship between predictive variance (uncertainty) and segmentation accuracy (Dice coefficient) in terms of a) the segmentation model, and b) the number of training images. The two plots correspond to the box plots in Fig. 3 and Table 7, respectively. Each point represents a single muscle/bone (right and left sides combined). All experiments were performed at DB#1. $\rho$ indicates Pearson's correlation coefficient. Strong correlations were obtained between DC and the predictive uncertainty in all experiments.

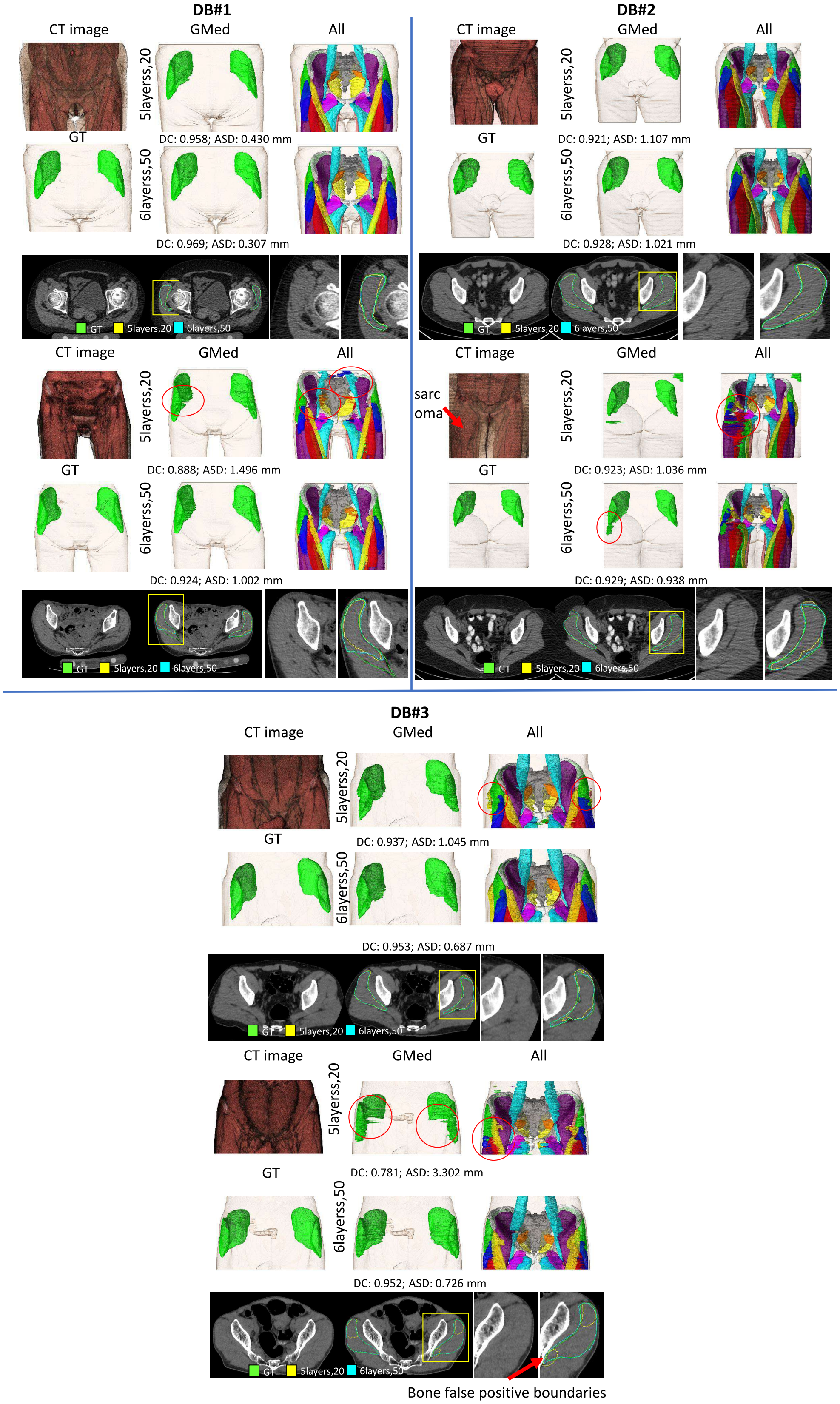

**Figure A.7.** Representative segmentation results from the validation databases DB#1-3, whose quantitative results are visualized in Fig. 6. For each database, the upper and lower cases correspond with the 5$^{th}$ (▲) and 95$^{th}$ (▼) quantiles of the predictive uncertainty. Red circles indicate improved locations using the 6layer,50 model. DC: Dice coefficient, ASD: Average symmetric surface distance.

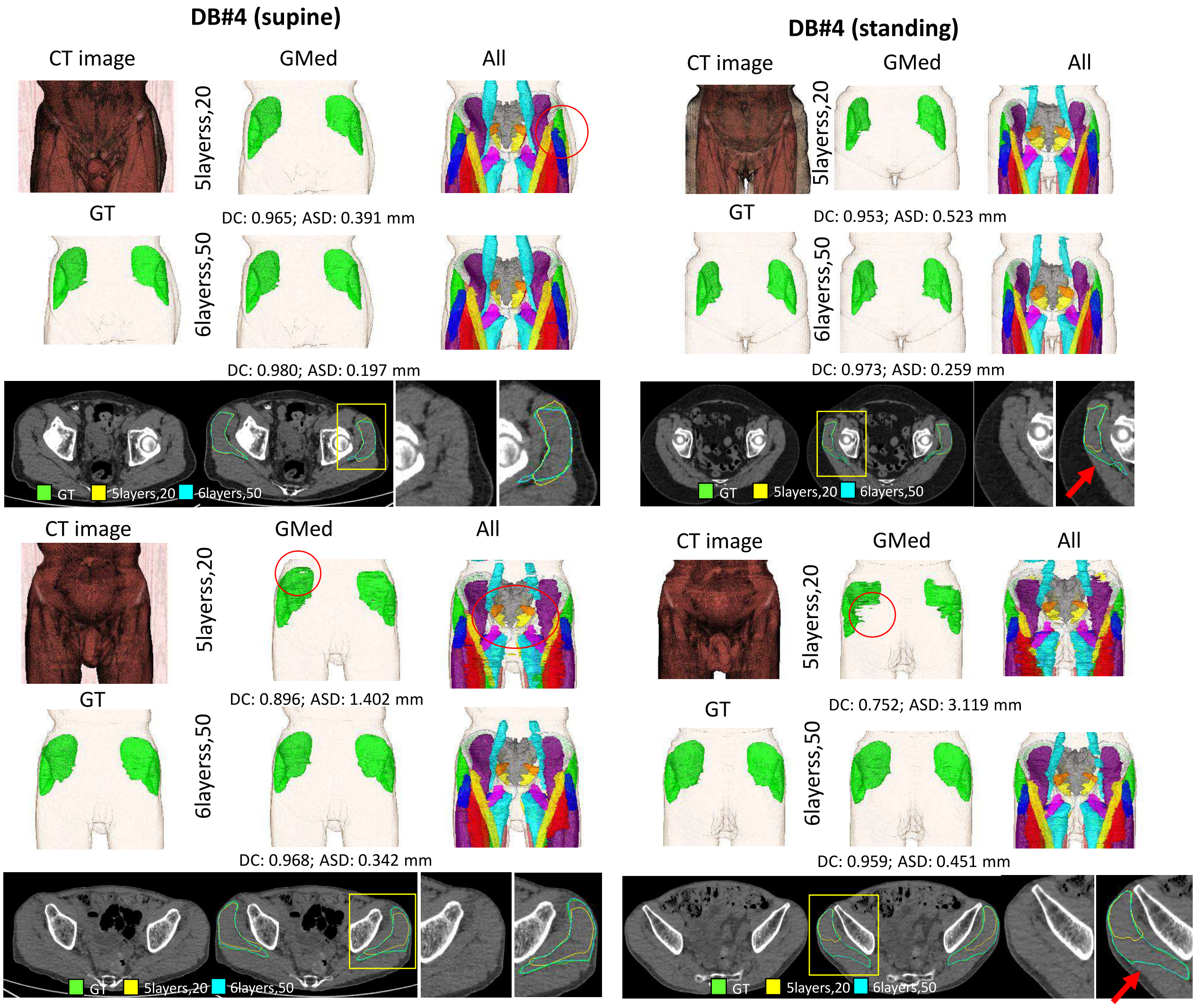

**Figure A.8.** Representative segmentation results from the validation databases DB#4 for the patient in supine and standing positionings, whose quantitative results are visualized in Fig. 6. For each database, the upper and lower cases correspond with the 5$^{th}$ (▲) and 95$^{th}$ (▼) quantiles of the predictive uncertainty. Red circles indicate improved locations using the 6layer,50 model. DC: Dice coefficient, ASD: Average symmetric surface distance.

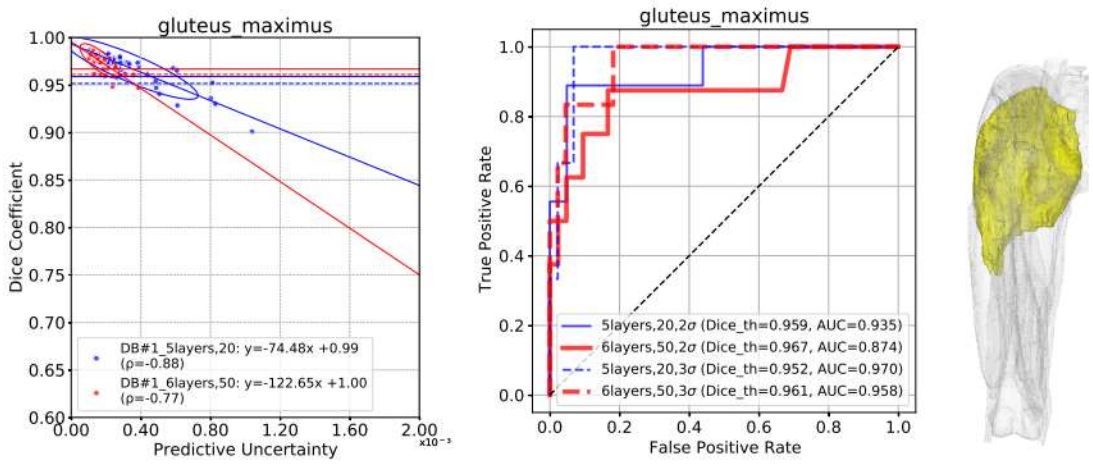

(a) Gluteus maximus

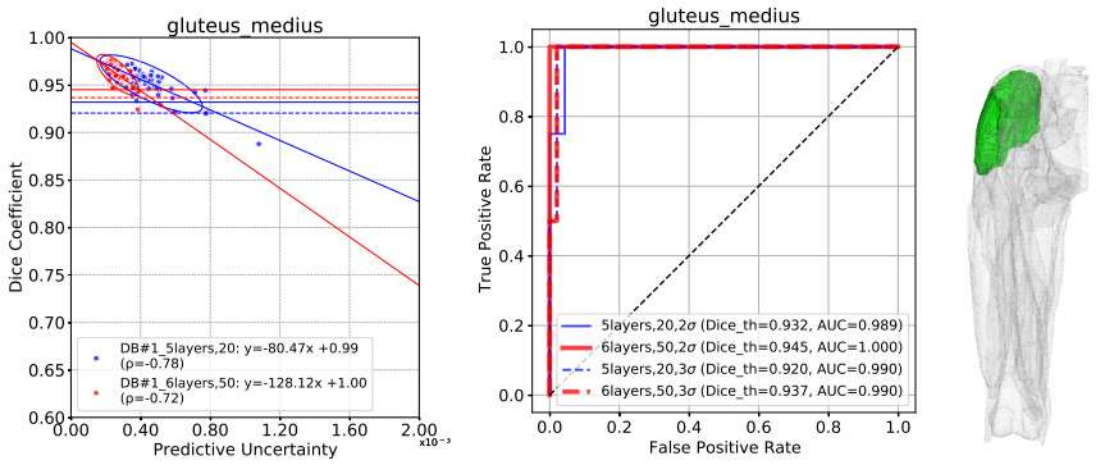

(b) Gluteus medius

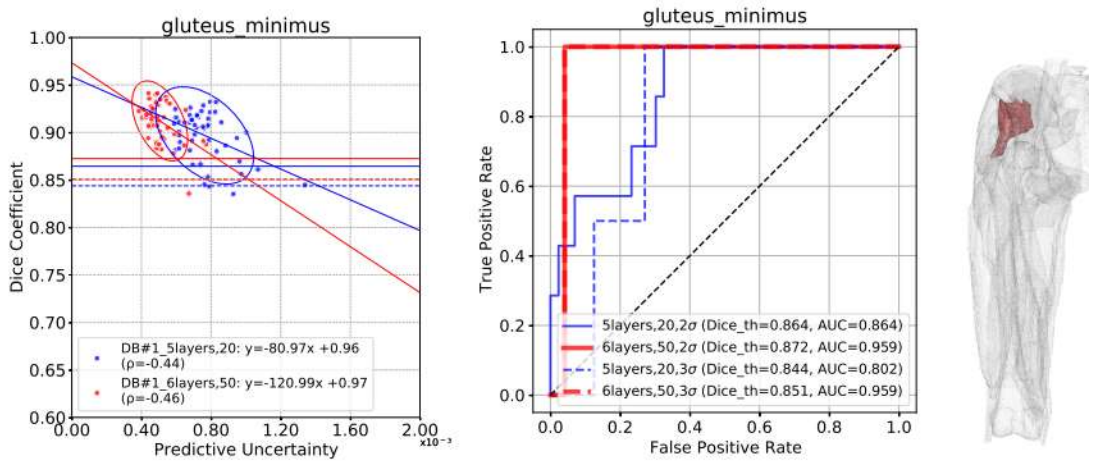

(c) Gluteus minimus

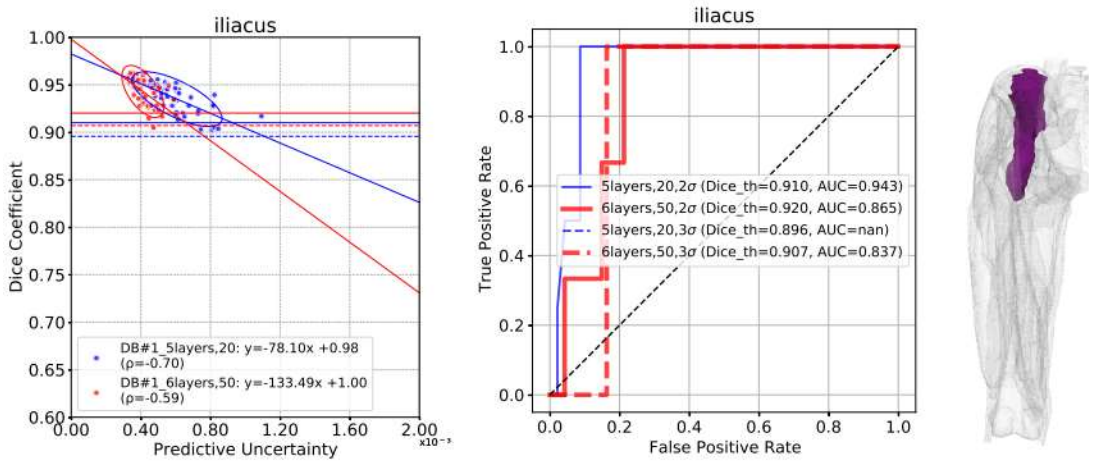

(d) Iliacus

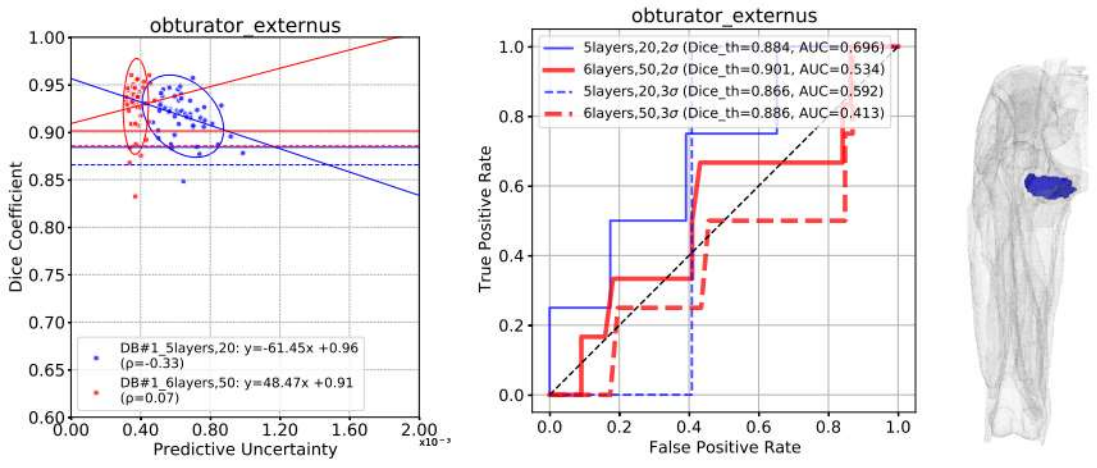

(e) Obtuator exturnus

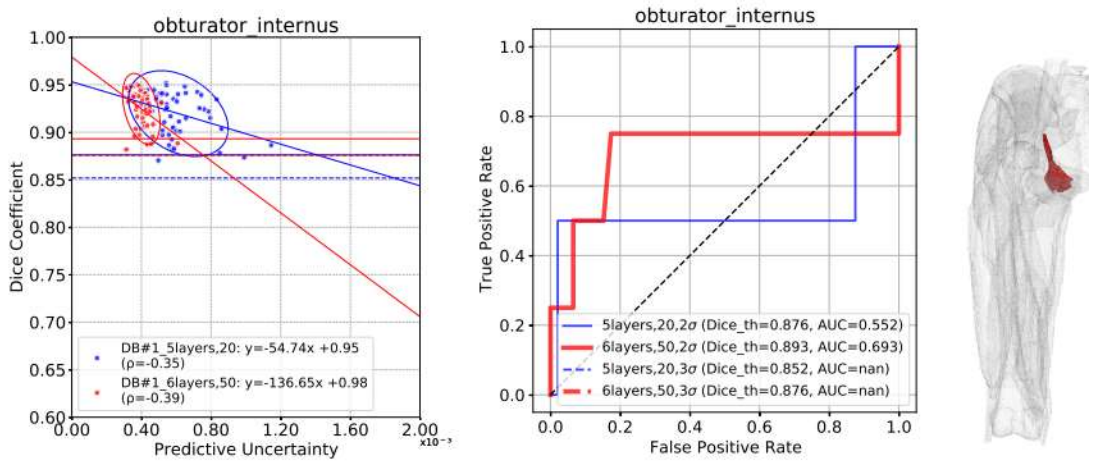

(f) Obtuator internus

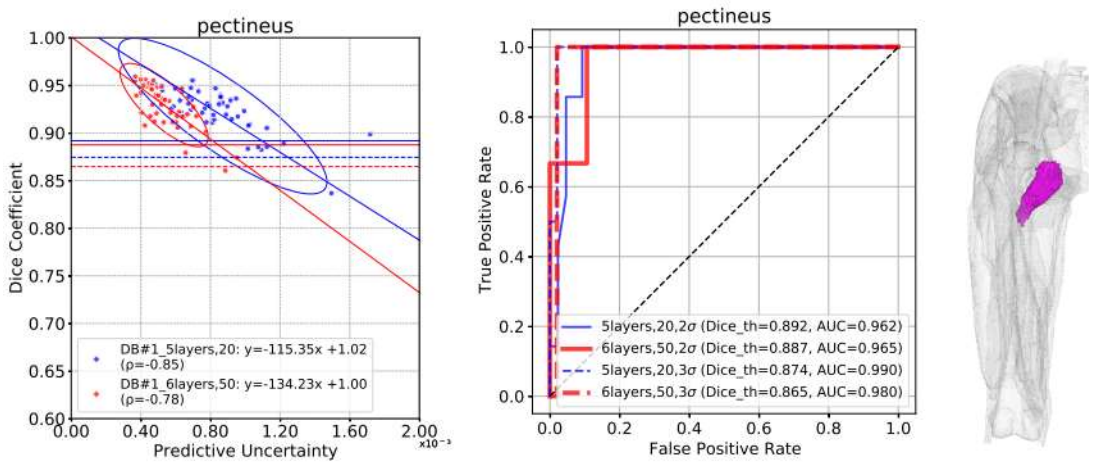

(g) Pectenius

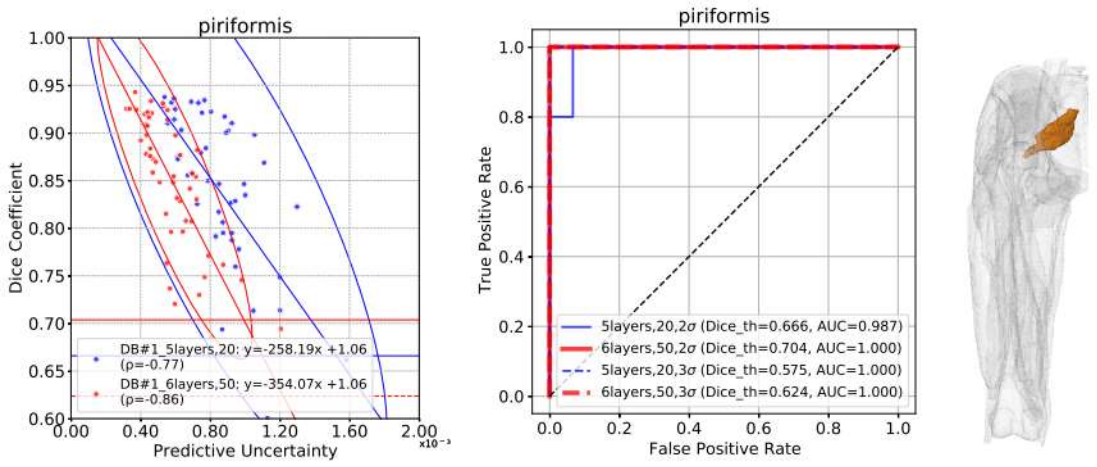

(h) Piriformis

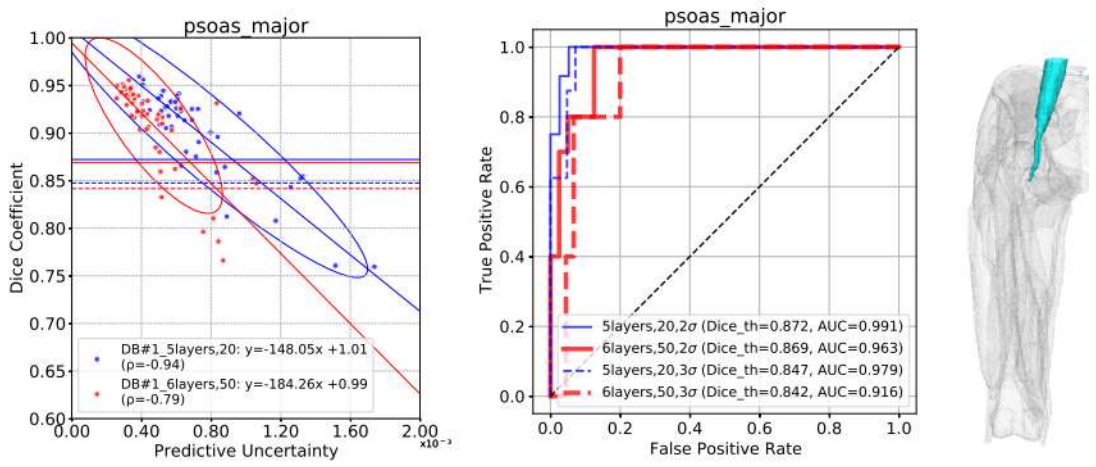

(i) Psoas major

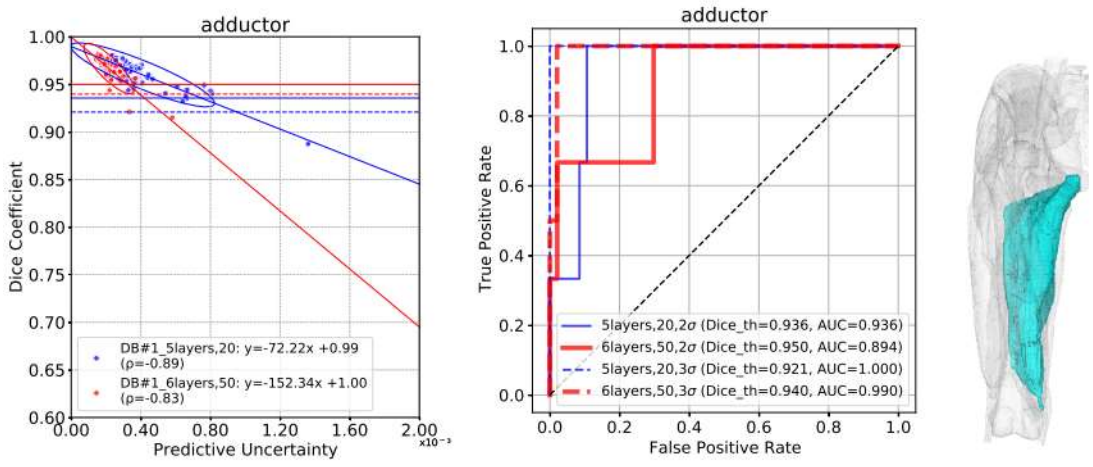

(j) Adductors

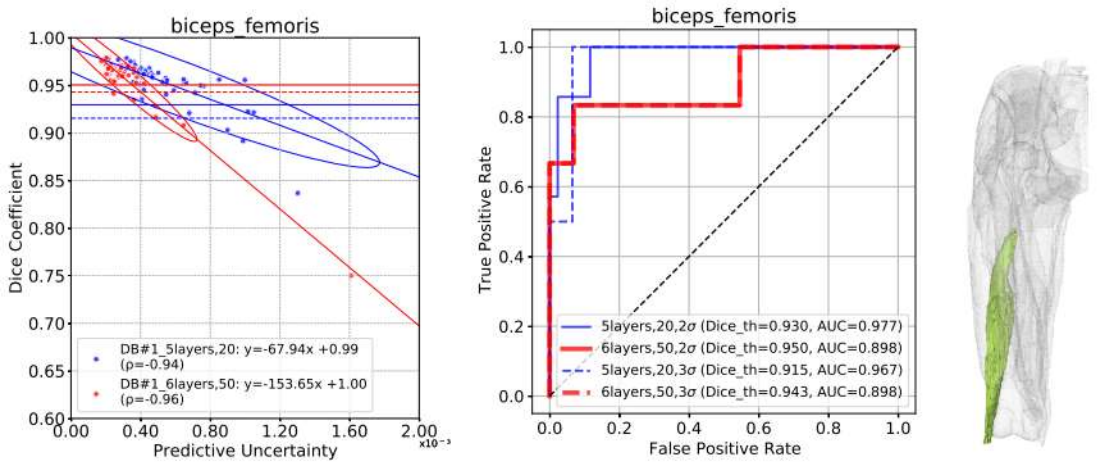

(k) Biceps femoris

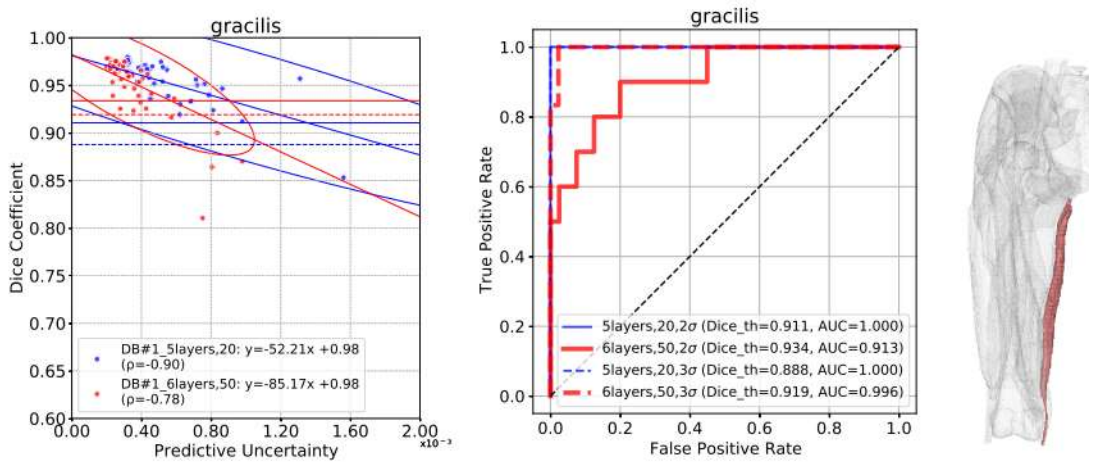

(l) Gracilis

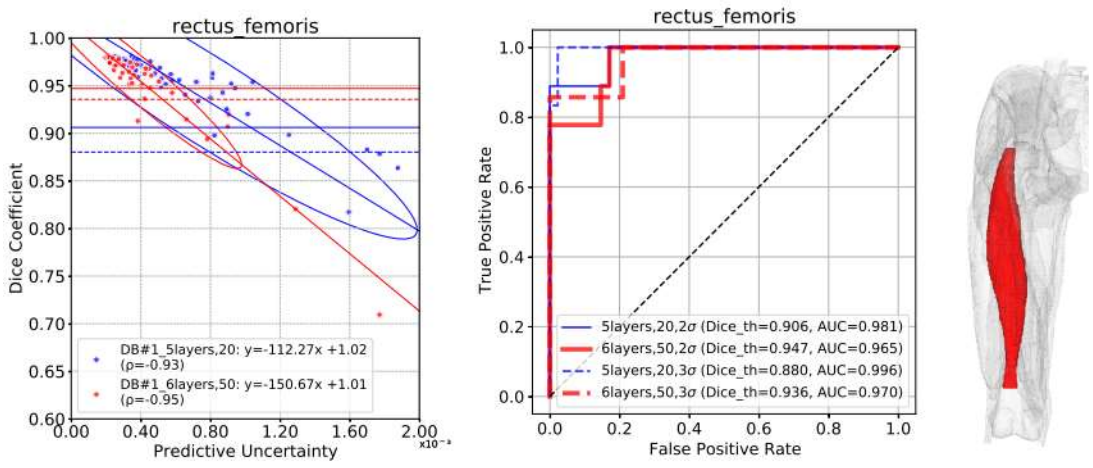

(m) Rectus femoris

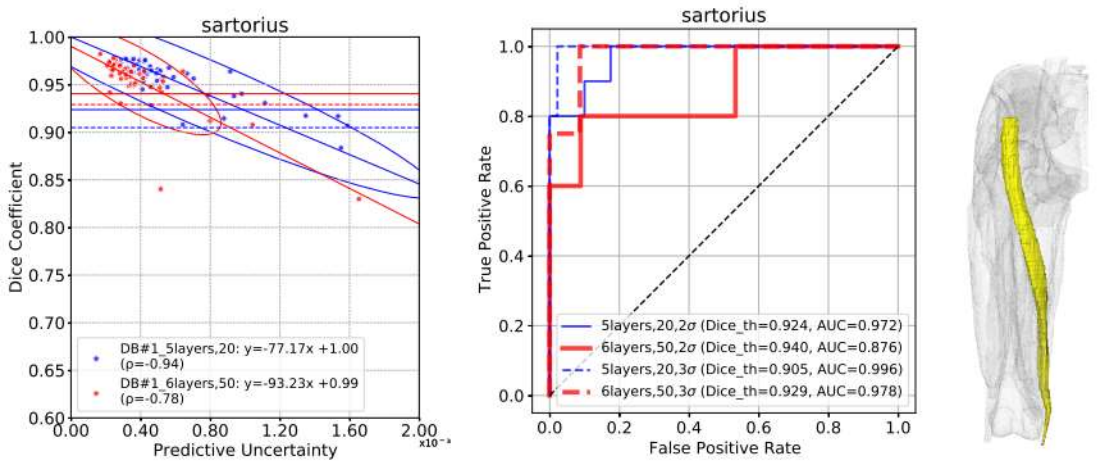

(n) Sartorius

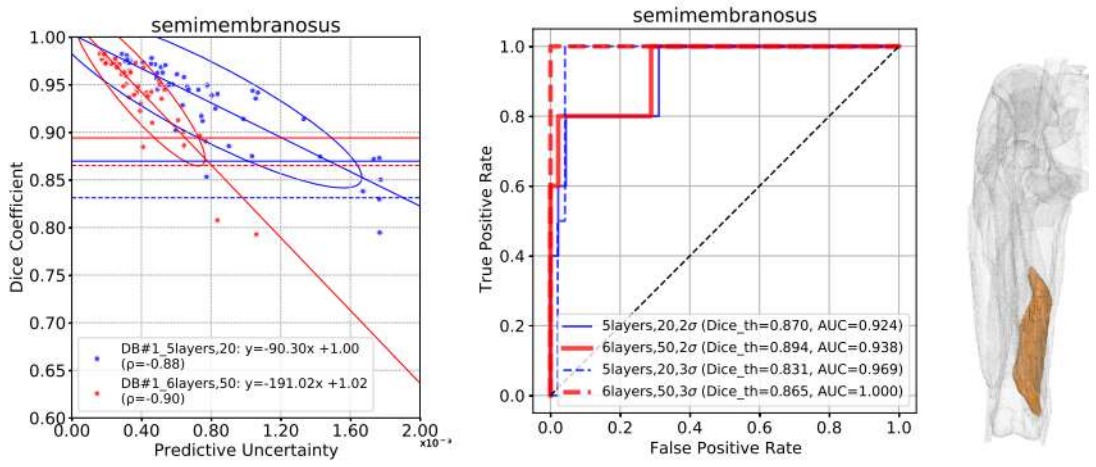

(o) Semimembranosus

contd.

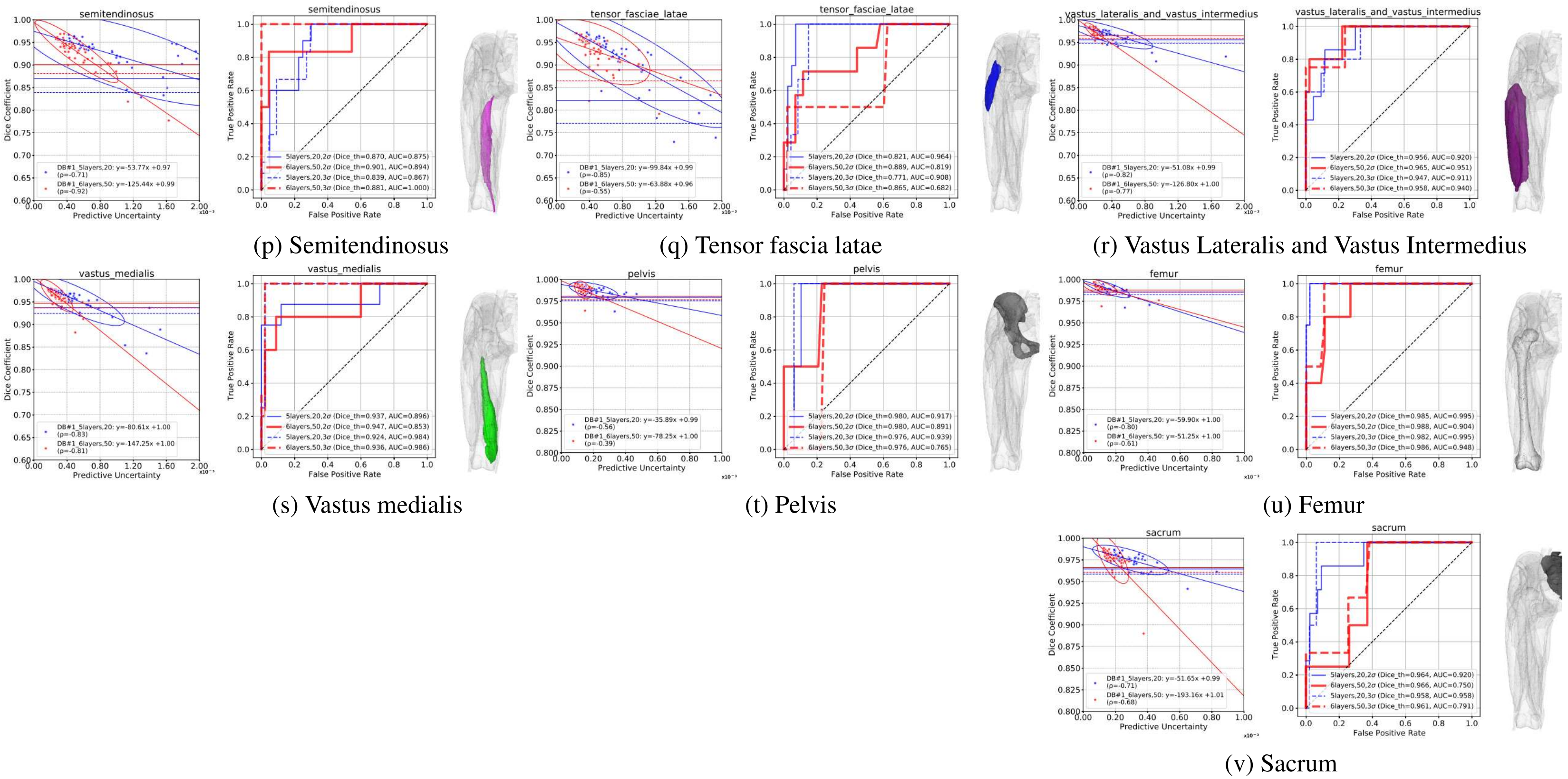

(p) Semitendinosus (q) Tensor fascia latae (r) Vastus Lateralis and Vastus Intermedius

(s) Vastus medialis (t) Pelvis (u) Femur

(v) Sacrum

**Figure A.9.** Relationship between the predictive uncertainty and segmentation accuracy (Dice coefficient; DC) in each structure at DB#1, with receiver operating characteristic (ROC) curves of detecting the inaccurate ($2\sigma$ of DC) and failed ($3\sigma$ of DC) based on the predictive uncertainty. In each plot, solid and dashed lines indicate the $2\sigma$ and $3\sigma$, respectively. Blue and red lines indicate the 5layers,20 and 6layers,50 models, respectively.

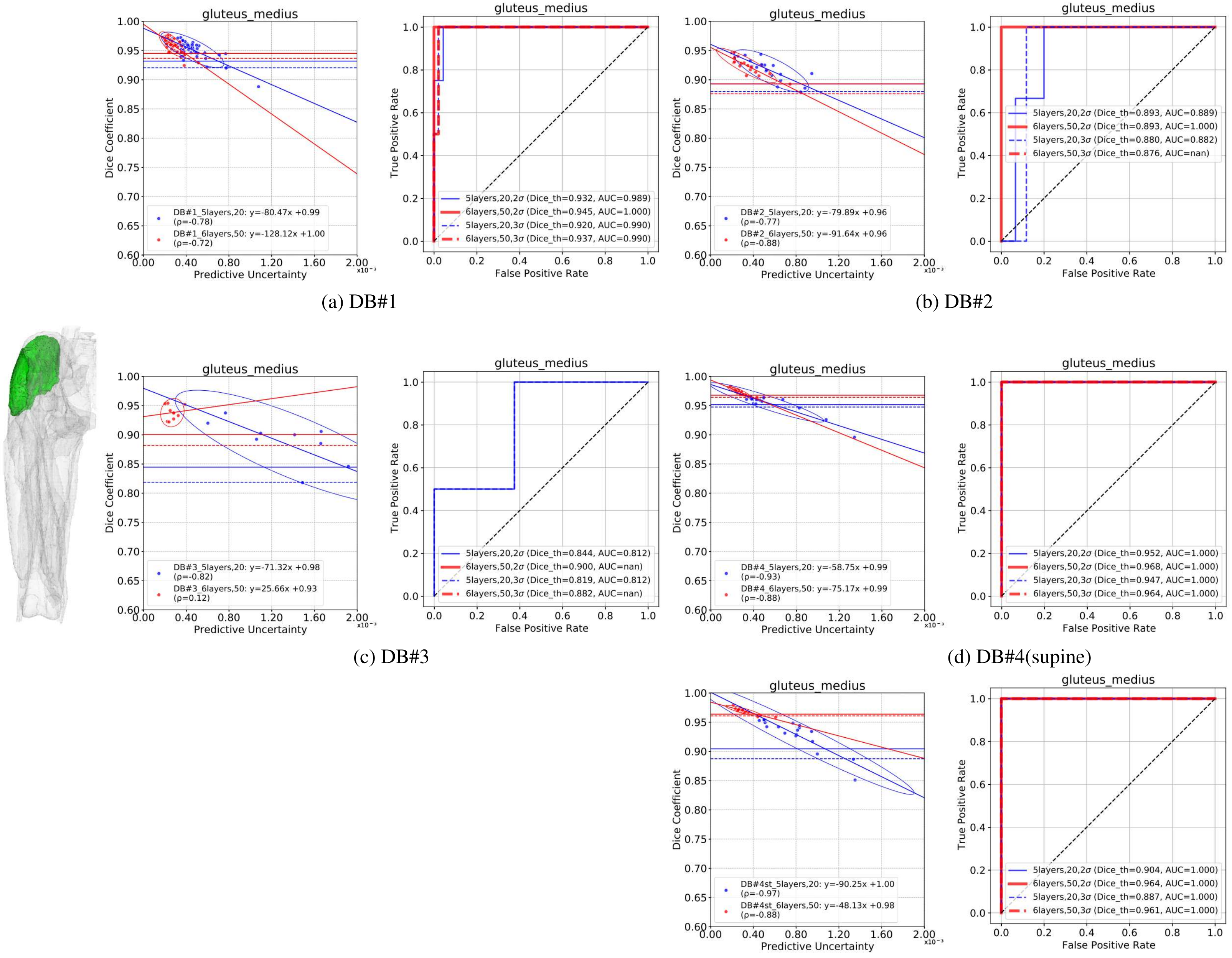

(a) DB#1

(b) DB#2

(c) DB#3

(d) DB#4(supine)

(e) DB#4(standing)

**Figure A.10.** Relationship between the predictive uncertainty and segmentation accuracy (Dice coefficient; DC) of the gluteus medius at the four databases, with receiver operating characteristic (ROC) curves of detecting the inaccurate (-2$\sigma$ of DC) and failed (-3$\sigma$ of DC). In each plot, solid and dashed lines indicate the -2$\sigma$ and -3$\sigma$, respectively. Blue and red lines indicate the *5layers,20* and *6layers,50* models, respectively.

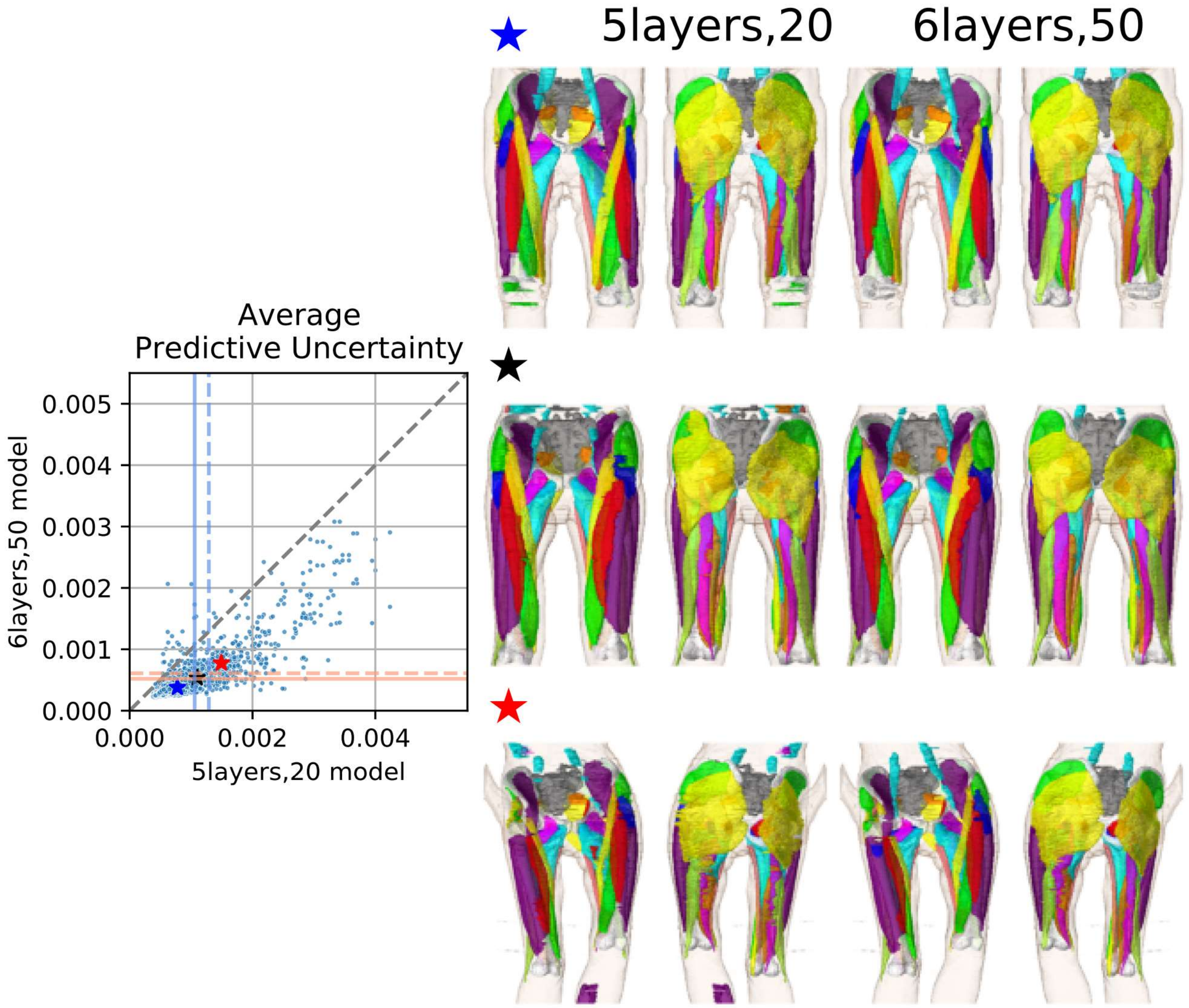

**Figure A.11.** Usability of the predictive uncertainty in detecting inaccurate and failed segmentations. The blue and orange lines correspond with the *5layers,20* and *6layers,50* thresholds, respectively. The solid and dashed lines correspond with the $2\sigma$ and $3\sigma$ thresholds, respectively, derived from DB#1 (See Fig. 4). The stars correspond with the representative cases based on the predictive uncertainty thresholds. ⋆ : $x<2\sigma$, ⋆ : $2\sigma<x<3\sigma$ ⋆ : $x>3\sigma$